\PassOptionsToPackage{dvipsnames,usenames}{xcolor}
\documentclass[conference]{IEEEtran}
\IEEEoverridecommandlockouts
\AtBeginDocument{%
  \providecommand\BibTeX{{%
    \normalfont B\kern-0.5em{\scshape i\kern-0.25em b}\kern-0.8em\TeX}}}

\PassOptionsToPackage{table}{xcolor}
\usepackage{cite}
\usepackage{amsmath,amssymb,amsfonts}
\usepackage{algorithmic}
\usepackage{graphicx}
\usepackage{textcomp}
\usepackage{balance}
\usepackage{todonotes}
\usepackage{acronym}
\usepackage{array}
\usepackage[inline]{enumitem}
\usepackage{listings}
\usepackage{cleveref}
\usepackage[dvipsnames]{xcolor}
\usepackage{xspace}
\usepackage[font=small,labelfont=bf]{caption}
\usepackage[htt]{hyphenat}
\usepackage{diagbox}
\newcolumntype{L}{>{\raggedright\arraybackslash}X} 
\usepackage{tabularx}
\usepackage{float}
\usepackage{multicol}
\usepackage{textcomp}
\usepackage{multirow}

\newacro{udf}[UDF]{User-Defined-Function}
\newacro{dbms}[DBMS]{Database Management Systems}
\newacro{cfg}[CFG]{Control Flow Graph}
\newacro{gnn}[GNN]{Graph Neural Network}
\newacro{dfg}[DFG]{Data Flow Graph}
\newacro{dag}[DAG]{Directed Acyclic Graph}
\newacro{mlp}[MLP]{Multi-Layer Perceptron}
\newacro{auc}[AuC]{Area-under-Curve}
\newacro{spaj}[SPAJ]{Sum-Project-Aggregate-Join}
\newacro{pl}[PL]{Programming Languages}
\newacro{ml}[ML]{Machine Learning}

\newcommand{\system}{GRACEFUL\xspace} 
\newcommand*\circles[1]{\raisebox{.5pt}{\textcircled{\raisebox{-.8pt}{#1}}}}

\begin{document}

\title{\system: A Learned Cost Estimator For UDFs\\
}

\author{\IEEEauthorblockN{
Johannes Wehrstein\IEEEauthorrefmark{1}, Tiemo Bang\textsuperscript{1}\IEEEauthorrefmark{2}, 
Roman Heinrich \IEEEauthorrefmark{1}\IEEEauthorrefmark{3}, 
Carsten Binnig\IEEEauthorrefmark{1}\IEEEauthorrefmark{3}
}
\IEEEauthorblockA{
\IEEEauthorrefmark{1} Technical University of Darmstadt
}
\IEEEauthorrefmark{2} Microsoft - Gray Systems Lab
\IEEEauthorrefmark{3} DFKI
}

\maketitle

\begin{abstract}
\acp{udf} are a pivotal feature in modern DBMS, enabling the extension of native DBMS functionality with custom logic. 
However, the integration of UDFs into query optimization processes poses significant challenges, primarily due to the difficulty of estimating UDF execution costs. 
Consequently, existing cost models in DBMS optimizers largely ignore UDFs or rely on static assumptions, resulting in suboptimal performance for queries involving UDFs. 
In this paper, we introduce \textbf{\system}, a novel learned cost model to make accurate cost predictions of query plans with UDFs enabling optimization decisions for UDFs in DBMS. 
For example, as we show in our evaluation, using our cost model, we can achieve 50$\times$ speedups through informed pull-up/push-down filter decisions of the UDF compared to the standard case where always a filter push-down is applied.
Additionally, we release a synthetic dataset of over 90,000 UDF queries to promote further research in this area.
\end{abstract}

\begingroup
\renewcommand\thefootnote{1}
\footnotetext{\vspace{-5ex}Work done while at UC Berkeley.}
\endgroup
\setcounter{footnote}{1}

\setlength{\skip\footins}{6pt}

\section{Introduction}\label{sec:introduction}
\textbf{Modern Databases Increasingly Face UDFs.} 
Rising data volumes and constantly evolving computing requirements increase the need to move computation as close as possible to the data. 
One important way to meet this demand in \ac{dbms} is to utilize \acfp{udf}---versatile stored procedures. \acp{udf} are crucial to encapsulate complex logic that cannot be efficiently expressed by traditional SQL constructs alone. 
In that way, they enable developers and data scientists to extend the native functionality of \ac{dbms} with custom operations tailored to specific application requirements. 
For example, \acp{udf} are frequently used for formatting values using string operations, custom aggregation operators, or complex conversion or filtering functions (e.g. conversion from Celsius to Fahrenheit or formatting timestamps). 
Therefore UDFs have emerged as a pivotal feature in modern \ac{dbms} and are executed billions of times daily in data centers \cite{DBLP:journals/pvldb/GuptaR21,DBLP:journals/pvldb/RamachandraPEHG17}. 

\textbf{UDF Optimizations are Crucial.}
The integration of \acp{udf} into database systems plays a key role in achieving fast and efficient query execution. 
Crucial for efficient execution is the query optimizer, responsible for selecting an efficient execution plan. 
However, \acp{udf} pose significant challenges for query optimization, as their impact on plan runtime is often poorly understood. 
For instance, while predicate push-down is a common heuristic that typically improves performance, in fact predicates involving \acp{udf} contradict this heuristic. 
Since \acp{udf} can be computationally expensive as they can include loops or external function calls, deferring their execution to later plan stages might improve performance.

\begin{figure}
    \centering
    \includegraphics[width=0.92\linewidth]{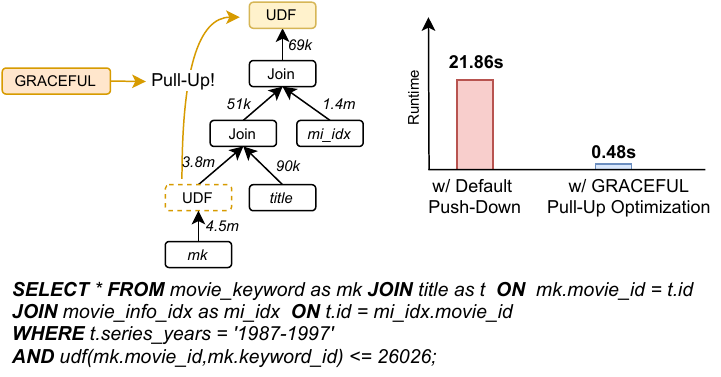}
    \vspace{-1ex}
    \caption{The impact of pull-up optimization an a SQL query with a UDF: carefully chosen pull-ups of \ac{udf} filters in a query plan can drastically reduce query runtime. The numbers along the query edges annotate the intermediate cardinalities.}
    \vspace{-4ex}
    \label{fig:pull_push_query_plan}
\end{figure}

\textbf{Limited UDF Support in DBMS Cost Estimators.}
Considering UDFs in query optimization is a decades-old problem, already raised in the late 90s, e.g., by Stonebraker in \cite{DBLP:conf/sigmod/Stonebraker91}, which discusses query optimization with complex predicates. 
Since then, several approaches focused on optimizing UDF computations to increase their performance \cite{DBLP:journals/pvldb/RamachandraPEHG17, DBLP:conf/cidr/FranzAHGMP24}. 
However, the task of estimating their costs in order to optimize query plans without having them seen before has not yet been addressed. 
This leads query optimizers to often fail in plan selection and have them fall back to na\"{\i}ve solutions if UDFs are involved. 
To the best of our knowledge, all available commercial and non-commercial systems today either use a static constant for the UDF cost or ignore the UDF cost completely. 
This is because DBMSs largely rely on traditional cost functions handcrafted for a pre-defined set of well-studied DBMS operators\cite{DBLP:journals/pvldb/LeisGMBK015,DBLP:conf/sigmod/ZhangIM0GLFHPJ22}. 

\textbf{The Importance of Optimizing UDF Queries.}
The lack of precise cost estimates significantly degrades performance when queries include \acp{udf}. 
While basic optimizations, like applying static rule sets on the \ac{udf} operator, are applicable, more sophisticated cost-based optimizations are not supported, as they lack cost estimates for the UDF. 
However, making informed cost-based optimization decisions of SQL queries which include a \ac{udf} can make a 10x, 50x, or even larger difference on the total query runtime. 
This is shown in \Cref{fig:pull_push_query_plan} where pulling up the UDF-filter to the very top of the query plan (instead of pushing it down to the table) reduced the runtime from 21.86 to 0.48 seconds. 

Let us now look at the example in \Cref{fig:pull_push_query_plan} in more detail to understand where this effect comes from.
As mentioned before, textbook-style rule-based query optimization assumes that a push-down of filter operation is always beneficial\cite{DBLP:books/mg/Ramakrishnan98,DBLP:books/daglib/0010423}. 
Though this assumption holds for non-UDF filter predicates, as the filter predicate evaluation is not a dominating factor, this rule does not necessarily hold for complex predicates involving UDFs. 
As shown in \Cref{fig:pull_push_query_plan}, with the presence of \acp{udf}, it can be actually beneficial to pull up a UDF filter predicate if its computational cost can be reduced as the UDF needs to be applied to fewer rows (69k rows instead of 4.5m rows) as shown in the example. 

\textbf{Our Proposal: Learned Cost Estimator for UDFs.}
In recent years, machine learning has shown promising results for the task of cost estimation and query optimization, opening a new avenue of database components \cite{DBLP:conf/icde/LiangCXYCX024,DBLP:conf/sigmod/WuMLNNPSRNK24,DBLP:journals/pvldb/HilprechtB22, DBLP:journals/pvldb/ZhaoCSM22,DBLP:journals/pvldb/SunL19,DBLP:journals/pvldb/MarcusP19,DBLP:conf/icde/HeinrichBKL24, DBLP:journals/pvldb/MarcusNMZAKPT19,DBLP:journals/sigmod/MarcusNMTAK22,DBLP:journals/pvldb/ZhuCDCPWZ23,DBLP:journals/pacmmod/DoshiZJMHABF23}. 
These ML techniques for cost estimation prove to capture complex effects of data and workloads. Thus far, they do not capture the complexities of \acp{udf} in queries.
UDFs have inner structure themselves---e.g., loops, branches, library calls---calling for a new class of more cost estimation ML models.
We hence propose \system (\textbf{GRA}ph-based \textbf{C}ost \textbf{E}stimator \textbf{F}or \textbf{U}ser-defined \textbf{L}ogic), a learned cost model for query plans containing \acp{udf}.

In general, estimating the runtime of arbitrary program code is a hard problem. 
This stems from the fact that determining runtime in the most general sense is equivalent to the Halting Problem, the undecidable problem of determining the termination of a Turing machine. 
To solve this challenge and predict costs for \acp{udf} accurately, we utilize the following observations:
(1) UDFs typically come with code of manageable structure, as shown by \cite{DBLP:journals/pvldb/GuptaR21}. 
They are typically composed of a comparably simple control flow of a limited number of loops and branches and have a limited size of only dozens to hundreds of operations. 
(2) In contrast to execution code outside a query plan, for \acp{udf} inside a query plan we have additional information on the data the UDFs are applied. 
This allows reasoning on how often a certain operation in the UDF will be executed or for how many rows an if/else branch condition will evaluate to true. 

\textbf{The Need to Generalize to Unseen UDFs.}
Although UDFs share structural similarities, the actual code can vary widely. 
Due to this diversity among \acp{udf}, it is crucial that the model can generalize across \acp{udf} and predict runtimes accurately for unseen elements. 
To address this, with \system we provide a learned cost estimation model that aims to capture the general runtime complexity of UDFs based on additional statistics on how often certain code paths are executed, which we derive from the base tables.
As we show in our paper, our model can thus
generalize out-of-the-box to (1) unseen \acp{udf} (unseen program code), (2) unseen SQL workloads (unseen SQL query patterns), and (3) unseen datasets. 

We achieve this by pre-training the cost model on a wide spectrum of queries with synthesized UDFs across various datasets.
Furthermore, we introduce a new representation for the UDF code based on the control flow graph of the UDF. 
With this representation, we train a \ac{gnn} model to produce a joint embedding of the UDF and query plan and predict costs for \acp{udf} with varying complexities, including loops, branches, arithmetic as well as string operators, and even library calls. 
Further, based on \system, we show that a pull-up advisor can be realized that operates on top of our cost model, achieving very robust and close to optimal query runtimes, highlighting the benefits of accurate \ac{udf} cost estimates for optimizing query plans.

\textbf{Adaptive Execution is Not to be Preferred.}
An alternative route to our UDF cost estimator is adaptive execution. 
With adaptive execution, the DBMS monitors the execution of the operators and would, in flight, e.g., depending on observed intermediate cardinalities, reconsider optimization decisions made so far, like the join order. 
This allows the DBMS to achieve fast execution times without needing a classical cost estimator since it will adapt the query plan during the execution.
However, we'd like to highlight that the adaptive execution of queries is highly complicated when integrated into a DBMS. 
It would require a completely different execution paradigm, which is, in most cases, incompatible with DBMS. 
Since cost models suit today's architecture, achieving the same speedups with a cost model is to be preferred. 
We, therefore, focus on non-adaptive systems in this work.

\textbf{Contributions.}
The main contributions of this paper are summarized as follows:

\begin{enumerate}[topsep=0pt, leftmargin=11pt] 
\item[\textbf{(1)}] We introduce \system, a \ac{gnn} based cost estimator capable of accurately predicting query runtimes for User-Defined Functions (UDFs). 
\item[\textbf{(2)}] We propose a novel transferrable representation for UDFs.
It captures the structure and input data distribution of UDFs and includes a novel method to leverage database statistics to estimate hit-ratios of UDF branches. 
This enables accurate cost estimation for previously unseen UDFs and databases.

\item[\textbf{(3)}] Our evaluation of \system demonstrates superior accuracy and significant end-to-end impact when optimizing pull-up / push-down for UDF queries.  

\item[\textbf{(4)}] Finally, we facilitate further research by releasing our code and benchmark dataset, comprising over 90,000 queries across 20 different databases. \footnote{\vspace{-5ex}https://github.com/DataManagementLab/Graceful.git}
\end{enumerate}
\textbf{Outline.}
The remainder of this paper is structured as follows: We first provide an overview of our approach in \Cref{sec:overview}. 
Next, we present our model's details, including the representation of UDFs (\Cref{sec:representation}) and statistics annotations required for precise cost estimation (\Cref{sec:card_annotation}). 
Further, we describe the joint representation of UDF and query (\Cref{sec:join_repr}) and details of the model training and inference (\Cref{sec:inference_training}) followed by the pull-up advisor \Cref{sec:pull_push}, which builds on our cost model.
Afterward, we discuss our novel UDF benchmark in \Cref{sec:udf_generation}, which we used for evaluation.
Finally, we evaluate both the cost-estimator as well as the pull-up advisor in \Cref{sec:evaluation} using this benchmark and conclude the paper with related work (\Cref{sec:related_work}) and a summary (\Cref{sec:conclusion}).

\begin{figure*}[ht]
    \centering
    
    \includegraphics[width=0.97\textwidth]{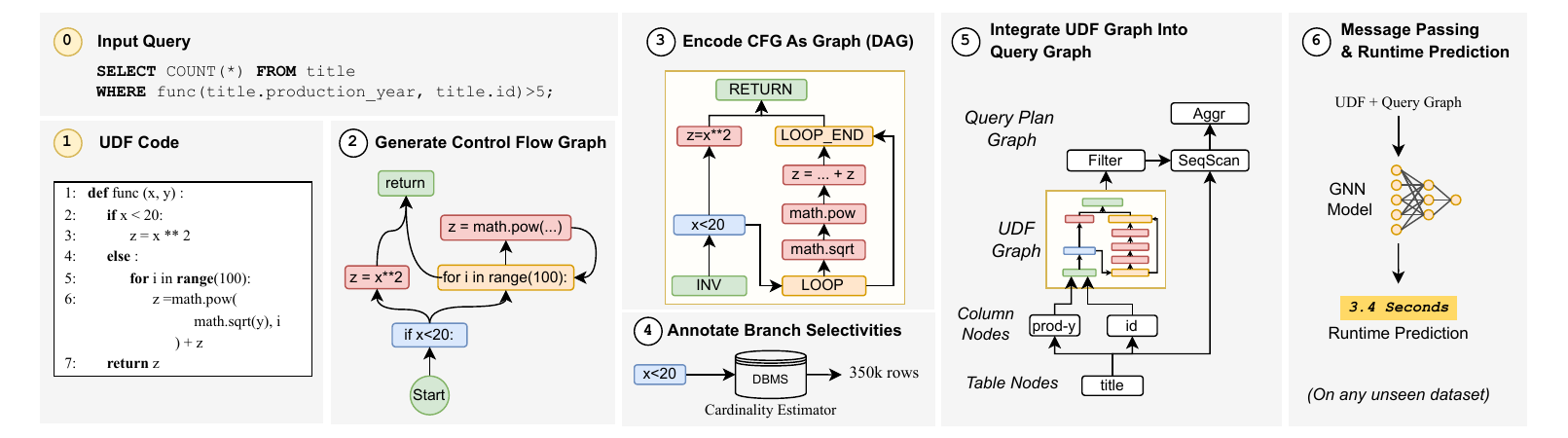}
    \vspace{-1ex}
    \caption{Cost Estimation Overview for a SQL Query \circles{0} with a UDF \circles{1}: For the UDF code, the CFG is computed \circles{2} and used as a transformed DAG representation of the code \circles{3}. On top of the CFG, we annotate estimates of the hit ratios of branches using a cardinality estimator \circles{4}. The UDF graph is combined with the query plan graph \circles{5}, and a GNN model predicts the runtime for the joint graph \circles{6}.}
    \label{fig:model_overview}
    \vspace{-3.5ex}
\end{figure*}\section{Overview of \system} 
\label{sec:overview}
In this work, we introduce \system, a novel learned cost estimator designed to predict the runtime of SQL queries which include UDFs as shown in \Cref{fig:model_overview}. 

\vspace{-.5ex}
\subsection{Key Ideas of Cost Model}

To facilitate cost estimation, we propose a novel representation that abstracts UDFs, SQL queries, and the data they are executed on in a graph structure. 
For this, we leverage the common structure of UDFs by describing them as \ac{cfg} as shown in \Cref{fig:model_overview} \circles{2}.
In a nutshell, the na\"ive \ac{cfg} represents code blocks as nodes and control paths as edges, forming the core of our UDF representation. 

\textbf{Our UDF Representation.}

An important aspect is that we perform multiple transformations and data-flow annotations on the \ac{cfg} as shown in \Cref{fig:model_overview} \circles{3} and \circles{4} to learn and predict UDF execution costs. 
Without this additional information, learning runtime estimates from the \ac{cfg} is as hard as learning runtime of general programs. 
For example, as part of the transformation, we add additional nodes (e.g., explicit loop ends) and transform the CFG into a single-statement CFG (split blocks of code into single lines).

Moreover, cardinalities need to be included in our representation, as they are critical for the task of cost estimation. 
This is essential for both query plan operators and the data flow inside UDFs. 
While the DBMS typically provides cardinality estimates for query plan operators, no equivalent exists for UDFs. 
To address this, we introduce a novel hit-ratio estimator that transforms UDF branch conditions into SQL queries and utilizes the DBMS's cardinality estimator to obtain predictions. 
The presence of database statistics in DBMS systems, and therefore the availability of knowledge about the data processed by a UDF, gives us a unique advantage over program codes in general. 

\textbf{Joint Query-UDF Representation.} 
For the cost estimation, we then embed the UDF graph jointly with the query graph and train a GNN model that predicts the runtime of the query plan, including the UDF runtime as shown in  \Cref{fig:model_overview}  \circles{5} and \circles{6}. 
For the query plan, we use a representation as proposed in our prior work \cite{DBLP:journals/pvldb/HilprechtB22}. 
As we have shown in \cite{DBLP:journals/pvldb/HilprechtB22}, annotating input and output cardinalities to each operator (e.g., filter, join) in the plan is crucial for accurate runtime estimations. 
However, a challenge for query plans with UDFs is that we cannot always derive an operator's output cardinality. 
For example, as shown in \Cref{fig:model_overview}, when a filter operation uses a UDF, the output cardinality cannot be estimated, as we further detail in \Cref{sec:pull_push}. 
As such, in this work, we develop a probabilistic approach inspired by Bayesian Networks and Regret-Optimization where the key idea is to iterate over the (unknown) UDF-filter selectivity and predict a distribution of the costs instead of just one cost estimate. 
With this distribution, query plan optimizations such as filter pull-up/push-down decisions can be enabled (cf. \Cref{sec:pull_push}).  

\textbf{Runtime Prediction.}
Based on the joined query-UDF graph a \ac{gnn} model is trained. 
The model involves a regression model as the final layer that consumes the graph representation learned by the GNN to predict the query's runtime. 
A core idea of \system is that it works in a zero-shot manner, i.e. providing out-of-the-box runtime prediction of UDF cost on unseen databases without the need to have seen the concrete UDF structure, dataset, or SQL query beforehand. 
This is enabled by using database-independent featurization, both on the UDF and the query nodes, as we discuss in \Cref{sec:design}.

\vspace{-.5ex}
\subsection{Discussion of Scope}\label{subsec:scope}
\vspace{-.5ex}

\textbf{Scalar Python UDFs.} In this work, we focus on scalar \acp{udf}, where an input is processed row-by-row, and a single output is returned for each input row. 
Each scalar \ac{udf} can contain diverse statements as supported by commercial DBMSs and include branches, loops, arithmetic, and string operations, as well as calls to external libraries (e.g., math or numpy).
In this paper, we focus also on Python UDFs.
However, it can be easily extended to other languages as our representation for UDFs (cf. \Cref{sec:design}) is language-independent. 
Moreover, we believe that our approach can also be extended to other types of UDFs like aggregate \acp{udf} \cite{DBLP:journals/pvldb/GuptaR21} e.g. by introducing additional node types describing the aggregation operation.

\textbf{Non-UDF Queries}. As real-world systems have to handle both queries with and without UDFs, we have designed and evaluated \system also for non-UDF queries. 
To achieve high accuracy for non-UDF queries, \system builds on the concepts of \cite{DBLP:conf/cidr/HilprechtB22} for the query-operators.

\section{\system Design}
\label{sec:design}

In the following, we explain our learned estimator for queries containing \acp{udf} (\Cref{fig:model_overview}). 
We first detail the graph constructions for representing the UDF-graph with its input features in \Cref{sec:representation}, followed by our UDF selectivity annotation strategy (\Cref{sec:card_annotation}). 
Afterwards, we explain the embedding of the UDF graph into the query graph in \Cref{sec:join_repr} and then discuss the model architecture, including training and inference of \system (\Cref{sec:inference_training}).

\vspace{-1ex}
\subsection{UDF Representation} \label{sec:representation}
\vspace{-.5ex}

In the following paragraphs, we will first outline how we represent the UDF code of a query as a graph-based model. 
For that purpose, we first explain how to derive a graph representation from the UDF source code. 
Then, we detail the node types and features of the model. 
Finally, we explain how UDF featurization and query plan information are connected. 

\textbf{Control-Flow Graph as Basis.}
The backbone of our graph representation is the \ac{cfg} of the UDF code~\cite{DBLP:journals/corr/abs-2208-07461}. 
The starting point for our graph structure is a standard representation of a CFG created with \textit{python\_graphs} \cite{DBLP:journals/corr/abs-2208-07461}. 
The \ac{cfg} comprises nodes for computations (basic blocks) and control flow statements like if/else conditions, loops, as well return or exception statements. 
The CFG connects all these nodes to a graph with edges indicating the possible execution paths of the UDF. 
For example, an if/else node has edges pointing to the basic blocks of either branch.
As an important first extension, we add nodes into the graph to model the data flow from input data into the graph. For this, we connect UDF nodes with column nodes (which represent the inputs of the \ac{udf})  from the query graph.
The column nodes contain important information such as data types for the UDF graph.

\textbf{Transforming the CFG.} Next, we transform the UDF graph to improve the learning efficiency of our model to enable topological learning (cf \cref{sec:inference_training}). 
For this, we derive an acyclic graph, replacing cyclic edges of loops with explicit loop nodes. 
For example, the cyclic dependency of the CFG in \Cref{fig:model_overview} \circles{2} is replaced with acyclic representation 

 in \circles{3}. 
More details on how we model loops are discussed also below. \textbf{Single-statement CFG.}
Moreover, as another important transformation, we break up complex computation statements as individual nodes in the graph for accurate representation. 
For that purpose, the basic blocks of the \ac{cfg} are split into multiple statements so that each block will be reduced to atomic statements (single statement \ac{cfg}).

For example, we split nested function calls into separate nodes as we illustrate for math function calls and arithmetic operations in \Cref{fig:model_overview}\circles{3}.

However, arithmetic operations that occur in the same statement (i.e., line of code) are not split up but kept in a single node, to reduce the overall graph size, which we have seen beneficial in our experiments. 
This trades off the accuracy and efficiency of the representation.

\textbf{Handling Loops}
One important aspect of our UDF graph is how we model loops. 
To model a loop, we use an acyclic structure as shown in \Cref{fig:model_overview} \circles{3} with a \texttt{LOOP} node, which indicates a loop start. 
Furthermore, we set a \texttt{loop\_part} flag on all the operators (i.e., nodes) inside a loop. 
However, with an increasing number of operations between the loop's start node and the last node in the loop, the propagation of the loop information in our \ac{gnn} model diminishes. 
To address this, we explicitly introduce a \texttt{LOOP\_END} node, which is usually not part of a \ac{cfg}, explicitly marking the end of the loop. 
To enhance the \ac{gnn}s capability for such long loops, we, in addition, introduce a directed connection between \texttt{LOOP} and \texttt{LOOP\_END} nodes. 
That way, the \texttt{LOOP\_END} node aggregates also representations of code segments that appear before the loop with the code in the loop and synthesizes them with the loop's characteristics. 
This is important as in our graph all information is aggregated in the root node of a UDF (i.e., the return node). 
The additional connection between a \texttt{LOOP} and \texttt{LOOP\_END} thus helps fusing loop-related information (type of the loop, code execution inside the loop) and pre-loop computations in the overall embedding of the UDF.

\textbf{Various UDF Node Types.}
To represent \ac{udf} code as a graph, we introduce five types of nodes, representing the main execution-cost-relevant building blocks of a \ac{udf}. 
These five node types can be divided into two categories: 
The first category of nodes are types that model control flow and computations. 
These include node types for the computations (\texttt{COMP}), for/while loops (\texttt{LOOP}), and branches within the \ac{udf} (\texttt{BRANCH}). 
The second category addresses the overhead of the \ac{udf} for invocation and returning of the processed values, i.e. the transformation of the input and output data, and includes the node types \texttt{INV}(Invocation) and \texttt{RET}(Return). \begin{table}
    \centering
    \scalebox{0.88}{
    
    \begin{tabular}{llll}
Node Type & Feature & Description & Type \\ \hline
\texttt{\textcolor{ForestGreen}{INV}} & \texttt{in\_rows} & Num incoming rows & int \\
 & \texttt{in\_dts} & Data types of UDF args & cat \\
 & \texttt{nr\_params} & Num of UDF args & int \\ \hline
\texttt{\textcolor{BrickRed}{COMP}} & \texttt{in\_rows} & Num incoming rows & int \\
 & \texttt{lib} & Lib call used & cat \\
 & \texttt{ops} & Arithmetic ops used & cat \\
 & \texttt{loop\_part} & If node is part of a loop & bool \\ \hline
\texttt{\textcolor{NavyBlue}{BRANCH}} & \texttt{in\_rows} & Num of incoming rows & int \\
 & \texttt{cmops} & Cmp op in branch cond & cat \\
 & \texttt{loop\_part} & If node is part of a loop & bool \\ \hline
\multirow{2}{*}{\begin{tabular}[c]{@{}l@{}}\texttt{\textcolor{brown}{LOOP /}} \\ \texttt{\textcolor{brown}{LOOP\_END}}\end{tabular}} & \texttt{in\_rows} & Num of incoming rows & int \\
 & \texttt{loop\_type} & loop type used (for/while) & cat \\
 & \texttt{nr\_iter} & Num of loop iterations & int \\
 & \texttt{loop\_part} & if node is part of a loop & bool \\ \hline
\texttt{\textcolor{ForestGreen}{RET}} & \texttt{out\_dts} & Data type of UDF output & cat \\
 & \texttt{out\_rows} & Num outgoing rows & int
\end{tabular}
    }
    \vspace{-1ex}
    \caption{Features of the \ac{udf} node types. The featurization is transferable and can be applied to unseen code.}
    \label{tab:node_features}
    \vspace{-7ex}
\end{table}

\textbf{Transferable Featurization.}
UDFs appear in a very large variety - probably any UDF appearing in the wild in DBMS is unique. 
Therefore, we cannot expect to have them seen all beforehand, requiring a transferable representation that can be applied to any (unseen) UDF. 
To provide a representation of code components within the UDF such that they can generalize to unseen code structures, we propose transferable features as outlined in \Cref{tab:node_features}. 
The idea is that the features capture the general code complexity of different node types. 
An important feature of all nodes is the number of incoming rows, which is important for accurate runtime estimates, particularly for loops, or to differentiate which branch of an if/else condition will be executed how often (cf. \Cref{sec:card_annotation}). 
To ensure the transferability of the computations performed in the \acp{udf}, we assume a superset of arithmetic and string operations and library calls, covering all major usages or arithmetic and string operations as well as numpy and math library calls, and encode them in one-hot vectors. 
Similarly, remote calls could be modelled in future work.\footnote{{This could also be achieved by specific remote/network nodes. However, both require a deterministic runtime not dominated too much by varying network latency or other unpredictable external factors.}}
For branches we abstain from encoding the concrete comparison literal, since such would not be transferable. 
In contrast, by annotating cardinalities, as described before, and encoding the categorical comparison operator ($<,>,!=$,...), we achieve transferability and still cover the characteristics that are affecting the execution-cost of the branch. 
Similarly, for the \texttt{INV} node, we abstract from the names of the input arguments and instead featurize the number of UDF arguments. 
In addition, their data types are represented in a vector, where each position corresponds to a python dtype and the number of arguments with this dtype is stored in this position. 
Similarly, for the \texttt{RET} node the output datatype is featurized.
This enables to capture important information about the data processed and overheads connected to data-type conversions. 

\vspace{-.5ex}
\subsection{UDF Selectivity Annotation}
\label{sec:card_annotation}
\vspace{-.5ex}

In this section, we discuss our method of annotating input rows for the different UDF nodes, raising a novel hit-ratio estimation that uses the DBMS cardinality estimator.

\textbf{Hit-Ratios Estimation / Branch Prediction.}

UDFs frequently come with complex control flow, including loops and branches\cite{DBLP:journals/pvldb/GuptaR21} which cause tuples to take alternate paths through the same UDFs.

In \Cref{fig:model_overview} \circles{2}, for example, a tuple with $x=1$ takes the left branch and incurs only the cost of a single arithmetic operation. 
In contrast, a tuple with $x=42$ takes the right branch and incurs the cost of many loop iterations.Important is that the cost of a tuple depends on its exact path. 
As such, estimating how many tuples hit individual branches is imperative based on the distribution of input tuples.

To estimate frequencies of code paths, we introduce a hit-ratio estimator that leverages the database statistics to estimate by which percentage a given branch will be used for this query. 
\textbf{Using the DBMS Cardinality Estimator.}
We introduce a hit-ratio estimator for UDFs branches by combining code analysis and off-the-shelf cardinality estimation. We first trace the conditions for all paths through the UDF.For each path, we then retrieve all conditions occurring along the path (i.e. if/else conditions) and rewrite them (together with the joins and filters applied before the UDF) into an SQL query: 

\vspace{-.5ex}
\begin{lstlisting}[language=SQL, breaklines=true, basicstyle=\footnotesize]
SELECT * FROM tables WHERE joins_before_UDF AND filters_before_UDF AND branch_conds_inside_UDF;
\end{lstlisting}
\vspace{-1ex}

For this query, we estimate the output cardinality which in fact is the branch frequency.
The generated queries are fairly simple, and our \ac{dag} representation ensures a finite number of these.\footnote{{For this step, we excluded the residual edges between LOOP\_START and LOOP\_END node.}}  
\vspace{-.5ex}
\subsection{Joint Query-UDF Representation.}\label{sec:join_repr}
\vspace{-.5ex}
Finally, to learn an end-to-end embedding for end-to-end cost estimation of a SQL query that includes a UDF, we embed the \ac{udf} graph into the surrounding query plan that determines other important factors, such as the join order, which we get from the DBMS optimizer.An example for the joint graph is illustrated in \Cref{fig:model_overview}\circles{5}. 
The query plan graph is an extension of~\cite{DBLP:journals/pvldb/HilprechtB22} as we explain next.The integration of the UDF graph into the query plan graph is executed through a series of directed edges. 
These edges establish connections between nodes of the UDF graph and the surrounding query plan graph and vice-versa. 
For this, we integrate parts of the \ac{dfg} into our graph representation. 
First, we add additional information about the data flow related to the input columns of the UDF, i.e., modeling which columns are used in which computations, enriching the graph with essential computational dependencies. 
\texttt{COLUMN} nodes in the query graph are linked to the invocation (\texttt{INV}) node of the UDF graph when used as an input of the UDF. 
Similarly, \texttt{COMP} nodes operating directly on an input column of the UDF are connected. 
This ensures that the model nodes of the UDF acquire detailed information about input data types as well as invocation overhead, such as data conversion. 

Conversely, we model the output of the UDF with

edges from the return (\texttt{RET}) node of the UDF graph back to the subsequent query operators.

For example, we create edges to \texttt{FILTER} nodes when using the UDF in a filter predicate or \texttt{OUTPUT COLUMN} nodes when using the UDF in an aggregation or projection. 
Notably, we explicitly annotate on the filter node whether it is operating on a \ac{udf} output by introducing a \textit{on-udf} feature. 
This showed to be beneficial in our experiments since it hints the model whether a filter is complex and probably expensive or rather simple.
\vspace{-1ex}
\subsection{Model Architecture}
\label{sec:inference_training}
\vspace{-.5ex}

To finally estimate the cost of a \ac{udf} in a query plan, we employ a \ac{gnn} architecture based on a GNN-MLP where each node type in our graph directly translates into a node type of the GNN and a final MLP produces the cost prediction based on the embedding the GNN produces. 
For this, our model leverages the inherent \ac{dag} structure of query execution plans in databases and in the UDF to apply a GNN with topological message passing. 
The GNN-MLP model consists of a node encoding phase, where the node features are embedded using an embedding layer, followed by a message-passing phase. 
Afterward, during the topological message passing phase, messages flow from the table/column nodes along the edges through the UDF graph and the query plan graph updating the node states. 
With message passing, the query operator and UDF embeddings are aggregated into one joint representation. 
This approach captures both the query plan's structural information and the code's details in the UDF. 
A regression model consuming the joint representation finally produces the cost estimation.

\vspace{-.5ex}
\section{Pull-up/Push-down Advisor}
\label{sec:pull_push}
\vspace{-.5ex}

In this section, we show how our model can be used for end-to-end query optimization. 
In particular, we present a solution to decide whether to pull-up or push-down filter operations with UDF predicates.
\vspace{-.5ex}
\subsection{Why this Problem?}
\vspace{-.5ex}

Ordering filters in the presence of UDFs
is a decades-old problem of high importance. 
Plans with good ordering offer significant speed up\cite{DBLP:conf/sigmod/Stonebraker91}.
These plans, however, have to be found under the absence of cardinality estimates after UDFs and unreliable cardinalities high in the plan.
As such, the main challenge is comparing plans with UDF push-down/pull-up alternatives under highly uncertain costs---despite an accurate cost model.

\begin{figure}
    \centering
    \vspace{-1ex}
    \includegraphics[width=.95\linewidth]{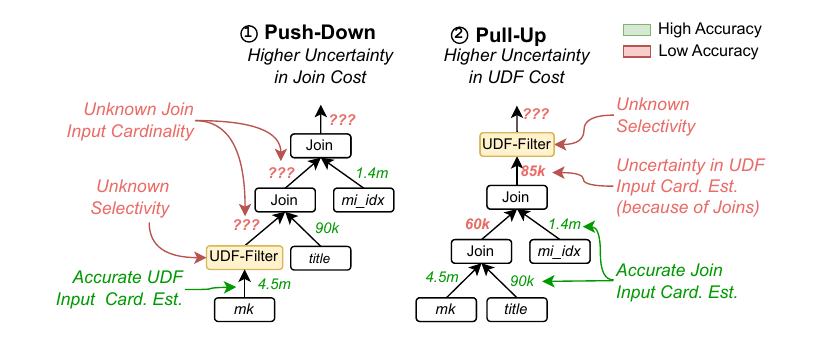}
    \vspace{-2ex}
    \caption{\circles{1} With the push-down of the UDF, the uncertainty in estimating the cost of join and other post-UDF operators increases drastically since the selectivity of the UDF-filter cannot be estimated and subsequent cardinality estimates are therefore inaccurate. \circles{2} Conversely, with a pull-up of the UDF, the UDF now has inaccurate cardinality estimates since, with the increasing number of joins, the accuracy of cardinality estimates drops leading to inaccurate cost estimates for the UDF.}
    \label{fig:pullpush_uncertainties}
    \vspace{-4ex}
\end{figure}

Hence, the query optimizers effectiveness in evaluating different plan alternatives when pulling up or pushing down a UDF filter, will be affected by the different levels of uncertainty in the cardinality estimations. 
We explain this problem in more detail with the simplified example of \Cref{fig:pullpush_uncertainties}: 
\begin{enumerate*}
    \item \emph{Uncertainty with Push-Down \circles{1}:} No cardinality estimator can predict accurate cardinalities after an UDF filter.
    For subsequent operators, only the UDF input cardinality can serve as rough cardinality estimate---a fixed upper bound.
    This hinders the selection of efficient physical operators or join orders, with compounding effects of deeper UDF push-down.
\item \emph{Uncertainty with Pull-Up \circles{2}:}
Pulling up UDFs introduces substantial uncertainties as well.
Pull-up exposes UDFs with decreasing accuracy of cardinality estimates higher in the query plan\cite{DBLP:journals/pvldb/LeeDNC23}. 
These inaccuracies can be orders of magnitudes, causing vast errors in the cost estimates.

\end{enumerate*}
In the following, we introduce a novel optimization technique that tackles this problem.

\begin{figure}
    \centering
    \vspace{-1.5ex}
    \includegraphics[width=\linewidth]{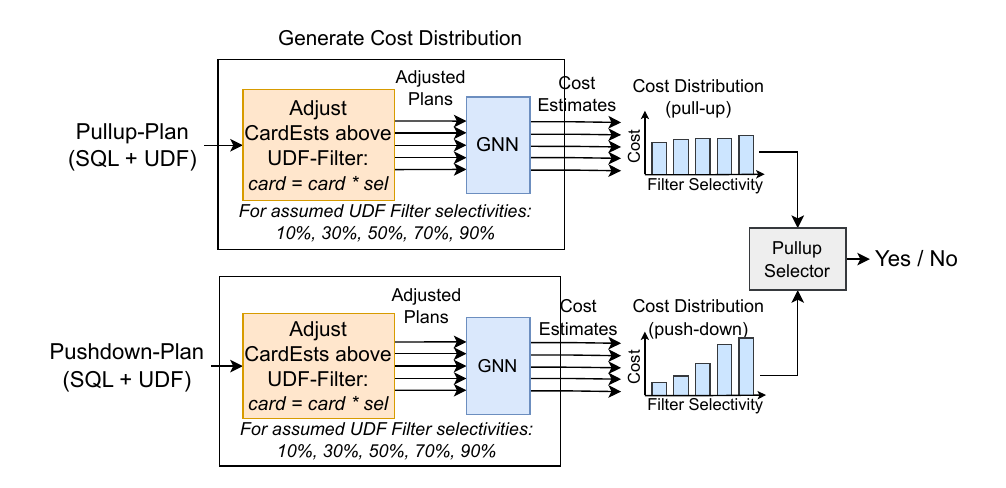}
    \vspace{-4ex}
    \caption{UDF Pull-up/Push-down Advisor: First, query plans are generated for both the pull-up and push-down plan. Next, \system enumerates multiple selectivity values for the (unknown and unpredictable) UDF-filter selectivity and creates a separate graph for each selectivity. These annotated plans are then passed through the GNN models, producing two cost distributions (one for the pull-up plan and one for the push-down plan). Based on a heuristics-based selection strategy we choose one plan alternative.}
    \label{fig:pullpush_architecture}
    \vspace{-4ex}
\end{figure}

\subsection{Regret Optimization}\label{subsec:pull_push_regret}

To address the problem of unknown post-UDF cardinalities, we now present a probabilistic approach inspired by Bayesian Networks and Regret-Optimization shown in  \Cref{fig:pullpush_architecture}. 
Here, the key idea is to iterate over a distribution of possible UDF-filter selectivities and predict the distribution of the costs of the entire plan for both the pull-up and push-down case. 
Heuristics are then employed to compare both distributions and decide on the pull-up / push-down as detailed in \Cref{sec:pull_push}.
To predict these cost distributions, we developed an approach where we iterate over the selectivity range going from zero, i.e. no rows returned by the filter, to one, i.e. every row is returned by the filter. 
For each selectivity level (i.e. $0.1, 0.3, 0.5, 0.7, 0.9)$ we produce a separate graph instance, where the cardinalities are adjusted to this assumed selectivity. This adjustment is done for all query plan operators on top of the UDF-filter by multiplying the assumed selectivity value with the annotated cardinality estimates.
By passing the generated pull-up and push-down graphs through our cost estimation model, we obtain the cost distribution across the full selectivity range.\vspace{-.5ex}
\subsection{Pull-up/Push-down Decision}

We have to decide between pull-up and push-down of UDF-filters without information about their selectivity. 
We hence consider the cost distribution of alternate plans over all possible selectivities.
In the simple case, one plan is clearly better due to lowest costs across all selectivities.

More likely, the cost of alternate push-down and pull-up plans are ambiguous, as illustrated in \Cref{fig:pullpush_architecture}. 
Either plan can have lowest for some selectivity range, e.g., the pull-up plan for low selectivities and the push-down plan for high selectivities. 
Crucially, also the distribution of costs can differ, e.g., the push-down plan can have a more extreme distribution with lower lows and higher highs.

We develop three heuristic-based decision strategies:
\begin{enumerate*}
    \item \emph{Upper Bound Cardinality (UBC):} We select the distribution where for the maximum selectivity of 1.0 (i.e., no rows are filtered out) the costs are lower. This gives us an upper-bound cost estimation.
    \item \emph{\ac{auc}:} We select the cost-distribution with the smaller area under the curve. This gives more accurate decision under the simplifying assumption that filter selectivity after the UDF is uniformly random.
    
    \item \emph{Conservative:} We only opt for a pull-up if it has strictly lower costs for all possible selectivities. That is, we keep the status quo of push-down, when it has lower cost for any selectivity. 
    
    This minimizes regressions in existing systems and always chooses not to pull up a filter if the two cost distributions overlap, and thus, there is a chance that pull-up a filter is more expensive.
\end{enumerate*}
These strategies provide different trade-offs between overall performance gain and risk of performance regression of individual queries.

The first strategy of upper-bound-cardinality has the highest risk of introducing regressions since it uses the assumption of a selectivity of 100\
On the contrary, the conservative approach minimizes the potential for regressions as discussed before. 
The \ac{auc} strategy is between those two strategies and tries to balance the gains of a pull-up with the risk of regression. 
In our evaluation in \Cref{sec:evaluation} we show that the conservative indeed provides the least regressions for a wide spectrum of workloads. However, other strategies are preferable if peak performance is important.

\section{A Novel UDF Benchmark}

\label{sec:udf_generation}

To the best of our knowledge, no benchmark is available that comprehensively evaluates how databases deal with \acp{udf}. 

Creating such a benchmark poses challenges, as datasets, queries, and \acp{udf} need should cover a diverse spectrum of realistic SQL queries with UDFs. 
This section addresses this gap by proposing a novel benchmark of 20 real-world databases and 90,000 UDF queries. 

\vspace{-.5ex}
\subsection{Benchmark Design}
\vspace{-.5ex}

Overall, a UDF benchmark should include (1) a diverse array of databases, (2) UDFs applicable to these databases, and (3) queries that invoke these UDFs. 
Our benchmark covers a wide range UDF complexities and mimics real-world use cases based on findings in \cite{DBLP:journals/pvldb/GuptaR21}. 
\Cref{table:udfbench} summarizes the characteristics of our benchmark.
\textbf{(1) Underlying Databases}.
Our benchmark provides 20 diverse databases, which are commonly used in works on zero-shot cost estimation\cite{DBLP:journals/pvldb/HilprechtB22, DBLP:conf/icde/LiangCXYCX024}. 
Eighteen of these are real-world datasets reflecting typical data distributions found in various domains, and overall three - SSB, TPC-H, and IMDB - are sourced from established benchmarks. 
These databases provide the semantic context for both UDFs and queries, as UDFs operate on the data stored within them, and queries are used for data retrieval. 

\textbf{(2) SQL Queries}.
The second component are the SQL queries that invoke scalar UDFs.
We cover cases where the UDF occurs as part of a filter predicate (i.e., filtering based on the output of a UDF) or as part of a projection or aggregation. 
For generating the surrounding SQL queries, we extend the methodologies of the workload generators proposed by Kipf et al. \cite{DBLP:conf/cidr/KipfKRLBK19} and Hilprecht et al. \cite{DBLP:journals/pvldb/HilprechtB22}. 
The generated SQL queries are SPJA queries which involve up to 5 joins and numerous filters and aggregations mimicking realistic workloads\cite{DBLP:journals/pvldb/RenenHPVDNLSKK24}. 

\textbf{(3) UDFs.}
The last and most difficult component are the UDFs themselves. 
Unlike the first two components, there is no ready-to-use corpus of \acp{udf} or an established methodology for generating UDFs. 
Therefore, we propose a novel approach to generate UDFs synthetically based on real-world UDFs. 
The UDFs produced by our generator reflect a range of complexities and structures commonly found in cloud databases\cite{DBLP:journals/pvldb/GuptaR21}, focusing on scalar functions that process single tuples.

\textbf{UDF Generation.}
We propose a novel process for the synthetic generation of UDFs based on an underlying database:
\begin{enumerate*}
\item \emph{Input Selection:} a table is randomly selected from the database, from which a set of columns is randomly drawn to serve as input for the UDF. 
\item \emph{\ac{udf} Structure Definition:} A high-level UDF structure is defined using the previously sampled input variables. 
This structure includes essential building blocks such as sequential computations, branches, and loops. 
For this, we are mimicking the structure and complexity of UDFs commonly used in cloud databases based on the findings of \cite{DBLP:journals/pvldb/GuptaR21}. 
According to their study of real-world \acp{udf} in production databases, we model the occurrence and complexity of these elements with loops occurring in 7\
\item \emph{Source Code Generation:} Based on the high-level structure, the final source code for the UDF is generated. 
Up to a hundred arithmetic and string operations will be generated in each UDF, in addition to calls to libraries like \textit{math} and \textit{numpy}.

\end{enumerate*}
\textbf{Generating UDF Code.}
Generating UDF code for benchmarking is challenging because the generated code must align with the database context. 
While syntactic correctness ensures that UDFs code compiles, semantic correctness is equally critical: the UDF logic must be compatible with underlying data. 
For instance, a UDF designed to format strings relies on specific characteristics of the input column, such as consistent data types or string patterns. 
Similarly, a UDF performing arithmetic operations must account for edge cases such as division by zero or operations on NULL values, which can cause runtime errors.

To address this, our approach flips the typical paradigm: instead of spending tremendous effort to generate \acp{udf} that conform to the data, we adapt the data to match the generated \acp{udf}. 
For this, after generating the \acp{udf}, we analyze their code for potential error cases and apply targeted data preparation steps. 
These include transformations like replacing NULL values with default substitutes, adjusting numerical ranges, or enforcing constraints to align with UDF logic. 
For instance, columns involved in conditional checks may be pruned or augmented to ensure that conditions within the UDF are valid.

By aligning the data with the \acp{udf}, we achieve semantic correctness without the need for painstakingly crafting \acp{udf} to suit the database. 
This process significantly reduces the number of runtime errors, enabling the efficient assembly of large-scale benchmarks while maintaining meaningful interactions between \acp{udf} and the data.

\vspace{-.5ex}
\subsection{A Resource for UDF Cost Models}
\vspace{-.5ex}
In addition to the queries and datasets,
our benchmark comprises the true runtimes of queries, including both push-down and pull-up plans. 
For collecting runtimes of the UDF queries, we execute them in DuckDB using CloudLab resources. 
As such, beyond its nature of a benchmark to stress the performance of UDF execution, it is an interesting resource on its own for the development of learned cost models with UDFs and can be used beyond the purpose of this paper.

\begin{table}[]
\scalebox{.88}{
\begin{tabular}{ll}
\textbf{Metric}                                                                    & \textbf{Values}                                                                                     \\ \hline
Number of Queries                                                                  & \begin{tabular}[c]{@{}l@{}}93.8k (72k w/ UDFs in filters,\\ 21k w/ UDFs in projection)\end{tabular} \\ \hline
Number of Databases                                                                & 20                                                                                                  \\ \hline
Total Runtime Of Benchmark                                                         & 142 hours                                                                                             \\ \hline
Query Complexity                                                                   & 1-5 Joins, 0-21 Filters                                                                             \\ \hline

\begin{tabular}[c]{@{}l@{}}
UDF\\
    - Number of Branches\\
    - Number of Loops\\
    - Number of Arithmetic / String Ops\\
    - Supported Libraries\\
    - Filter Selectivity\\
\end{tabular} & \begin{tabular}[c]{@{}l@{}}\\0-3\\0-3\\10-150\\Math, Numpy\\0.0001-1.0\end{tabular} \\
\end{tabular}
}
\vspace{-1ex}
\caption{Statistics of the created benchmark: both the UDFs and the filters are mimicking real-world use-cases\cite{DBLP:journals/pvldb/GuptaR21, DBLP:journals/pvldb/RenenHPVDNLSKK24}}
\label{table:udfbench}
\vspace{-5ex}
\end{table}
\section{Experimental Evaluation}
\label{sec:evaluation}
{\noindent
In this section, we evaluate the effectiveness and efficiency of our learned cost model with the following set of experiments:}
\begin{itemize}[leftmargin=*]
\vspace{-.4ex}
 \item \textbf{Exp 1: How accurate can \system predict UDF runtimes?} We assess the model's ability to predict the cost for SQL queries that invoke UDFs accurately. In particular, we showcase the ability of the model to generalize to unseen datasets and UDF structures.
 \item \textbf{Exp 2: How robust are the estimates across different UDF complexities?} Here, we test the model's performance across various types of UDFs that differ in complexity. This includes variations in the number of loops, branches, arithmetic operations, and the extent of library calls.
 \item \textbf{Exp 3: How does \system compare with other UDF Cost Estimation Approaches?} We compare our approach with other cost-estimation approaches that are able to deal with unseen UDFs and datasets. 
 \item \textbf{Exp 4: How do major design decisions of \system affect the prediction accuracy?} 
 We show an ablation study for the different design choices of our cost model and how the features that are used for the representation of the UDF influence the overall accuracy of the model.
 \item \textbf{Exp 5: What speedups does our pull-up advisor achieve?} Finally, we evaluate the performance and assess the robustness of our pull-up advisor, showing the potential of the cost model when integrated with query optimization.
\end{itemize}

\textbf{Experimental Setup.}\label{subsec:experimental_setup}
We conduct all our experiments using DuckDB 0.10.1 \cite{DBLP:conf/sigmod/RaasveldtM19} on an Intel E5-2660v2 (2.2 GHz, 2x10 cores) machine with 265 GB RAM running Ubuntu 22.04.
We use DuckDB since it allows a low overhead \acp{udf} execution, particularly when additional libraries (e.g. numpy) need to be loaded. 
Furthermore, for the evaluation of the model, we report the Q-Error, and for the last experiment, we report query runtimes. 
The Q-Error is defined as \ensuremath{Q = \max\left(\frac{\hat{y}}{\text{y}}, \frac{\text{y}}{\hat{y}}\right)} where $y$ is the predicted and $\hat{y}$ the actual value. As such, it measures the relative deviation between the estimated and true values.
The model is trained in a zero-shot manner on 90,000 UDF queries drawn from 20 datasets in our benchmark (\Cref{sec:udf_generation}). 
Of these, 19 datasets are used for training, while the remaining one is reserved for testing. 
While it is theoretically feasible to train the model in an instance-specific manner and tuning the model to the target dataset, this approach would likely yield even better results. 
However, we argue that such a training strategy incurs impractically high overhead, as it requires collecting a substantial number of UDF queries for each individual dataset.

\textbf{Accuracy on Non-UDF Queries.}
To enhance prediction accuracy for queries where joins and scans dominate runtime as well as maintaining high precision for non-UDF queries, we incorporate a small proportion of queries without \acp{udf} into the training data, constituting less than 10\% of the overall dataset.
Our experiments show that \system achieves competitive Q-Error for non-UDF queries with a median Q-Error of 1.21 and a 95th percentile of 2.02.
However we focus in the following experiments on queries with UDFs.

\textbf{Cardinality Annotation Methods.}
For runtime estimations cardinalities are crucial.
To investigate the robustness of \system against errors in cardinality estimation, we test \system with different state-of-the-art cardinality estimators. 
These include actual cardinalities (as upper baseline with perfect cardinalities), DeepDB \cite{DBLP:journals/pvldb/HilprechtSKMKB20} (a data-driven cardinality estimator, default configuration), WanderJoin \cite{DBLP:conf/sigmod/0001WYZ16} (a sampling-based approach, using 100 successful walks) and the estimates of the DuckDB optimizer.

\textbf{Baselines.}\label{subsec:baselines}
To evaluate the efficacy of our joint embedding and since there exists no UDF cost estimator currently, we compare \system with two split approaches where costs for query and UDF are predicated separately and added up.
For this, we isolate the UDF from the query graph and predict the UDF and query costs using separate models.
Other approaches cannot support UDFs on unseen databases\cite{boulos1997neural} or not at all \cite{DBLP:journals/pvldb/HilprechtB22,DBLP:conf/cidr/KipfKRLBK19,DBLP:conf/icde/LiangCXYCX024} and are therefore not included. 
To estimate the UDF costs we are using an extended version of the FlatVector approach \cite{DBLP:conf/icde/GanapathiKDWFJP09} (Flat+Graph) as well as the UDF part of our graph model (Graph+Graph).
We have chosen these approaches since they are easily applicable for UDFs only and have shown to be very competitive in general with other cost estimation approaches \cite{heinrich2025goodlearnedcostmodels}.
The Flat+Graph baseline is built on the FlatVector approach and represents a UDF with the number of loops, branches, and invocations of the different arithmetic, string, and library operations. 
Based on this feature vector, the per-tuple cost for the UDF are predicted and then scaled up to the estimated total number of rows processed by the UDF. 
To predict the per-tuple cost estimates from the FlatVector representation we employed XGBoost\cite{DBLP:conf/kdd/ChenG16}, a widely used, lightweight gradient-boosting model that has proven effective for FlatVector approaches\cite{DBLP:journals/pvldb/HilprechtB22} and is competitive with other cost estimation approaches \cite{heinrich2025goodlearnedcostmodels}.
For the Graph+Graph baseline we extract the UDF related part of \system to form a separate UDF cost estimator.
To predict the query-related costs in both baselines, we employed the query-related graph of \system.
To train the UDF- and query-models of the split baselines, we also split the training workload and trained the models separately, i.e., the UDF models only on UDF graphs and -runtimes and the query model on query plan and -runtime.

\vspace{-1ex}
\subsection*{Exp1: Accuracy across Unseen Databases}\label{subsec:exp1}
\vspace{-.5ex}
 \begin{table*}
 \centering
\scalebox{0.88}{
\begin{tabular}{l|l|rrr|lll|lll|lll|rr}
\textbf{Model} & \textbf{Card. Est.} & \multicolumn{3}{c|}{\textbf{Overall Error}} & \multicolumn{3}{c|}{\textbf{Pull-Up}} & \multicolumn{3}{c|}{\textbf{Intermediate Position}} & \multicolumn{3}{c|}{\textbf{Push-Down}} & \multicolumn{2}{c}{\textbf{Card. Est. Error}} \\ \cline{3-16} 
 & \textbf{Method} & \multicolumn{1}{c}{\textbf{Med.}} & \multicolumn{1}{c}{95th} & \multicolumn{1}{c|}{99th} & \multicolumn{1}{c}{\textbf{Med.}} & \multicolumn{1}{c}{95th} & \multicolumn{1}{c|}{99th} & \multicolumn{1}{c}{\textbf{Med.}} & \multicolumn{1}{c}{95th} & \multicolumn{1}{c|}{99th} & \multicolumn{1}{c}{\textbf{Med.}} & \multicolumn{1}{c}{95th} & \multicolumn{1}{c|}{99th} & \multicolumn{1}{c}{\textbf{Med.}} & \multicolumn{1}{c}{95th} \\ \hline
\system & Actual & \textbf{1.15} & 3.99 & 11.66 & \textbf{1.09} & 1.48 & 2.00 & \textbf{1.10} & 1.62 & 2.87 & \textbf{1.19} & 5.08 & 18.94 & - & - \\
\begin{tabular}[c]{@{}l@{}}Flat+Graph\end{tabular} & Actual & \textbf{1.71} & 7.88 & 33.14 & \textbf{1.67} & 6.97 & 29.08 &\textbf{1.70} & 7.16 & 28.85 & \textbf{1.71} & 7.94 & 33.35 & - & - \\
\begin{tabular}[c]{@{}l@{}}Graph+Graph\end{tabular} & Actual & \textbf{2.61} & 215.64 & 792.05 & \textbf{2.17} & 74.54 & 255.38 &\textbf{2.65} & 218.21 & 526.93 & \textbf{2.72} & 229.41 & 849.81 & - & - \\ \hline
\system & DeepDB \cite{DBLP:journals/pvldb/HilprechtSKMKB20} & \textbf{1.25} & 10.08 & 45.17 & \textbf{1.25} & 10.49 & 460.99 & \textbf{1.11} & 1.76 & 2.77 & \textbf{1.28} & 11.19 &44.15 & \textbf{1.47} & 247.08 \\
\system & WanderJoin \cite{DBLP:conf/sigmod/0001WYZ16} & \textbf{1.26} & 7.89 & 88.46 & \textbf{1.75} & 14.07 & 31.52 & \textbf{1.13} & 1.71 & 2.60 & \textbf{1.25} & 7.23 & 84.87 & \textbf{1.21} & 309.38 \\
\system & DuckDB & \textbf{3.32} & 30.14 & 84.70 & \textbf{3.08} & 40.42 & 132.48 & \textbf{2.25} & 24.89 & 177.52 & \textbf{3.50} & 28.76 & 80.30 & \textbf{6.29} & 528.43 \\ \hline
\end{tabular}
}

\vspace{-1.5ex}
 \caption{Cost Estimation Q-Errors of our model with different cardinality estimators for different positionings of the UDF in the plan. The models are trained on 19 datasets and tested on the 20th, and the errors are averaged across all 20 runs. The results of the model vary with precision of the cardinality estimates. However, with state-of-the-art cardinality estimation methods, our model can predict query runtimes accurately and provides robust performance indepedently of the positioning of the UDF in the query plan. Further, \system outperforms also the baselines \texttt{Flat+Graph} and \texttt{Graph+Graph}
 which shows the efficacy of our joint representation. 
 Cardinality estimation errors are denoted for the top nodes of the plans.
 }
 \label{tab:zs_cost_avg_datasets}
 \vspace{-5ex}
\end{table*}

A key property of \system is that it predicts precisely costs for unseen datasets that have not been part of the training data and thus can be used as an out-of-the-box cost estimator.
Therefore, in the first experiment, we assess the model's ability to accurately predict the cost for SQL queries that invoke UDFs on databases not previously encountered by the model. 
In particular, we employ a leave-one-out cross-validation approach over the 20 datasets of our benchmark, each containing approximately 4000 queries, and average the errors over the 20 runs. 
Furthermore, to show the effects of cardinality annotations, we evaluate \system with both perfect cardinalities and estimates of various methods. 

The results are summarized in \Cref{tab:zs_cost_avg_datasets}, showing high accuracies with a median Q-Error of 1.17 with actual cardinalities and 1.27 with estimated cardinalities from DeepDB\cite{DBLP:journals/pvldb/HilprechtSKMKB20}. 
In general, this experiment shows, that the predicted costs are highly precise across the board, independent of where the UDF is placed in the plan. 
However, the precision of the cost estimates is affected by the quality of cardinality estimates. 
Nevertheless, as we show, \system is robust against small errors in CE like those produced by DeepDB or WanderJoin. 
Comparing \system with the \texttt{Flat+Graph} baseline, our results show that this split approach utilizing FlatVector for the UDF even with actual cardinalities is clearly outperformed by \system.
In addition, both \system and \texttt{Flat+Graph} perform better than \texttt{Graph+Graph}, which is because the graph model is more complex and requires more training, that e.g. could be facilitated when trained jointly with the query part. 
Further, placing the UDF in the middle of the plan (\texttt{Intermediate Position}), not seen by the model during training, shows to be the sweet spot for cost estimation for \system since the number of operators performed after the UDF reduces (operations after the UDF have no accurate cardinality estimates), while the cardinality estimates of the UDF (num input tuples and inside the UDF) is still accurate because only few joins are executed before the UDF.

\begin{figure}
 \centering
 \vspace{-3ex}
 \includegraphics[width=\linewidth]{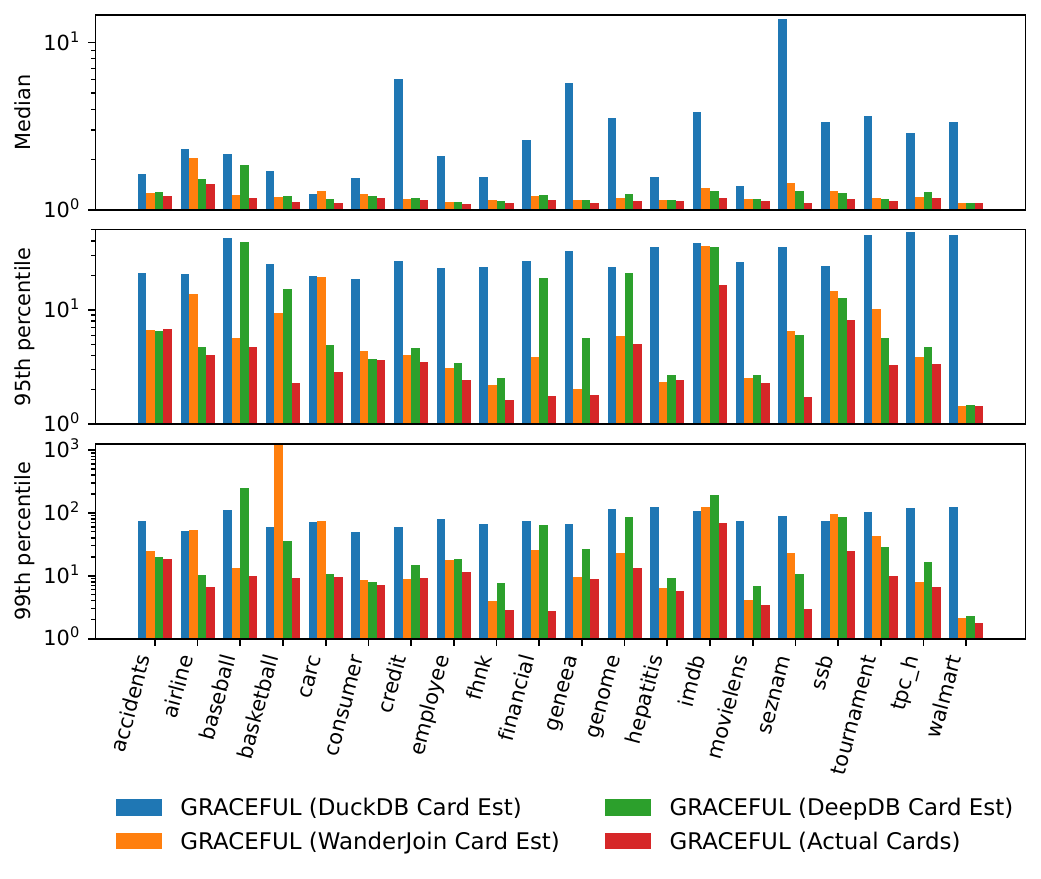}
 \vspace{-4ex}
 \caption{Cost Estimates for the different datasets. The errors are collected in a leave-on-out cross-validation fashion: the system is trained on 19 datasets and tested on the unseen 20th dataset. The y-axis shows the Q-Error (in logarithmic scale). The model shows a robust performance across all the twenty datasets. It is visible that some datasets are more challenging for some cardinality estimators, leading to higher errors in our cost estimates.}
 \label{fig:cost_est_per_dataset}
 \vspace{-2ex}
\end{figure}

\Cref{fig:cost_est_per_dataset} highlights {\system}s robust performance across various datasets, with median Q-Errors consistently below 1.3 for most datasets when using DeepDB's estimated cardinalities, except for the airline and baseball datasets, which present higher median Q-Error due to inaccuracies in the cardinality estimates. 
With actual cardinalities, even lower Q-Errors can be achieved across the board. 
This overall shows that our model consistently predicts precise \ac{udf} costs, which allows us to do plan optimizations on unseen databases. 
\vspace{-.5ex}
\subsection*{Exp 2: Zero-Shot Accuracy for Different Classes of UDFs}
\vspace{-.5ex}
\begin{figure}
 \centering
 \vspace{-.5ex}
 \includegraphics[width=\linewidth]{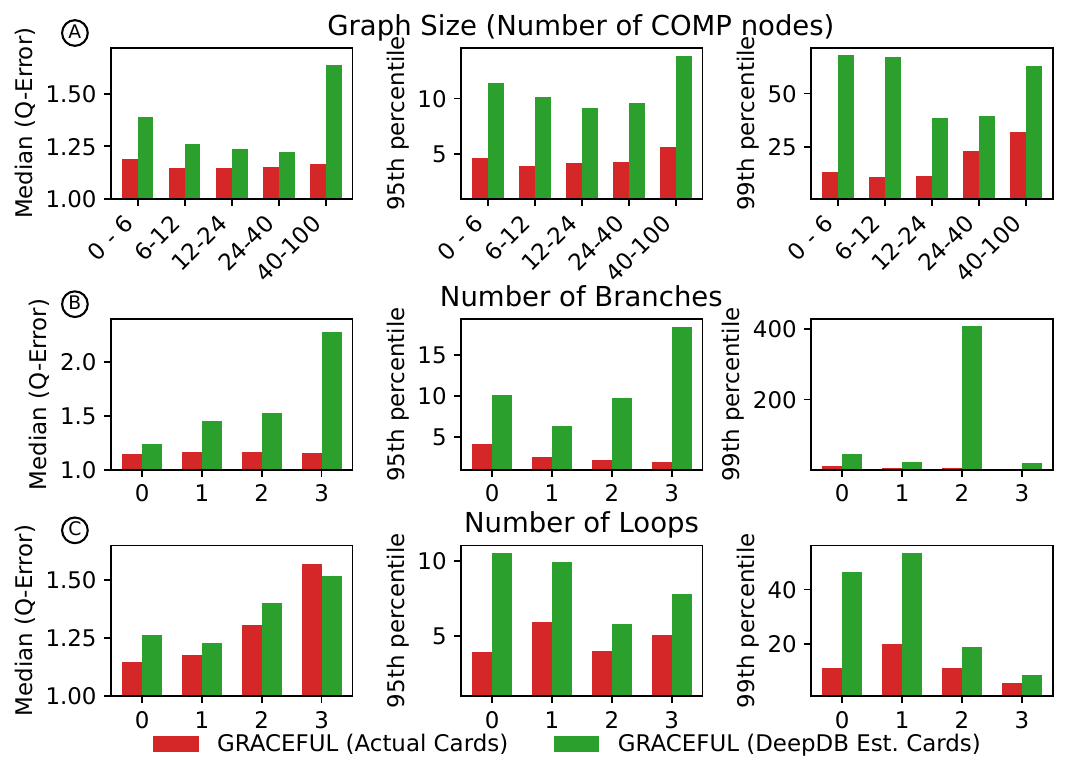}
 \vspace{-4ex}
 \caption{Q-Error distribution for an increasing number of branches and loops in the UDF and increasing graph size (nr of computations). The errors are shown for \system with actual cardinalities and estimated cardinalities using DeepDB. While the Q-Error with actual cardinalities stays constant for number of branches and graph size, the error with DeepDB annotations increases because of the errors introduced by the cardinality annotation method. This is a common problem with cardinality estimation. The higher percentiles show these trends less prominently, which is partly due to outliers in cost estimation of the surrounding query. 
 }
 \label{fig:error_over_multiplot}
 \vspace{-4ex}
\end{figure}

As DBMS are facing a broad variety of UDFs, we next analyze the model's performance for varying UDF complexities. 
We express the complexity by the three key aspects also discussed in \cite{DBLP:journals/pvldb/GuptaR21}: the number of operations in the \ac{udf}, the number of branches, and the number of loops. 
We now use both actual and DeepDB cardinality estimates for cardinality estimation as they have shown the best results in the previous experiment. 
The results are shown in \Cref{fig:error_over_multiplot}. 
Overall, these experiments indicate that the proposed model exhibits strong resilience to variations in UDF complexity, both in median cases and higher percentiles, showcasing precise predictions for various degrees of UDF complexity. 
In the following, we discuss the results along the different aspects.

\textbf{Varying Computations in UDF\circles{A}.}
First, we investigate how the model performs with UDFs of different code lengths, defined by the number of computations involved. 
The results demonstrate that the model scales effectively with increasing UDF size, showing only a marginal increase in the median Q-Error from 1.16 to 1.18 in the median case when using actual cardinalities. 
This indicates the model's robustness in handling larger UDF graphs, maintaining accuracy in cost estimation despite the growth in computation complexity. 
With estimated cardinalities, smaller error spikes are visible, while overall, we see a similar trend.

\textbf{Varying Number of Branches \circles{B}.}
The second experiment examines the model's performance with respect to the number of branches in the UDF, as they have a strong effect on \ac{udf} costs. 
Our findings reveal that while the model maintains high accuracy for actual cardinalities, as the number of branches increases, the performance degrades for DeepDB-estimated cardinalities. 
Specifically, the Q-Error rises to 2.3 when dealing with three branches, indicating challenges faced by cardinality estimators in accurately predicting the hit ratios of the branches.

\textbf{Varying Number of Loops \circles{C}.}
Finally, we assess the model's handling of UDFs with different numbers of loops. 
The analysis shows a slight increase in Q-Error, from a median value of 1.14 for UDFs without loops to 1.57 for UDFs with three loops. 
While we see this increase in the Q-Error here, we believe that this is no issue for practical usage since such loop-intensive UDFs are infrequent in production systems\cite{DBLP:journals/pvldb/GuptaR21}.
\vspace{-3ex}
\subsection*{Exp 3: Comparing With Other UDF Featurizations}\label{subsec:exp3}
\vspace{-.5ex}
To investigate the advantages of our graph-based UDF representation in contrast to a flat representation, we compare \system with the beforementioned FlatVector baseline on a UDF-only workload.
For this experiment, both models are trained again on a diverse set of 19 databases and then tested in a zero-shot fashion on unseen \acp{udf} from our benchmark on the 20th database. 
However, different from before, in this experiment, we restrict the workloads to simple queries without joins since, in these queries, the UDF cost dominate, and thus, the benefits of a graph-based representation vs. a flat representation of UDFs can be best shown.

\Cref{tab:flat_vector_comparison} shows the results, highlighting that the FlatVector approach achieves a median Q-Error of 1.89 (using actual cardinalities) and 2.01 (using DeepDB). 
In contrast, our trained model demonstrates superior performance with median Q-Errors of 1.29 (using actual cardinalities) and 1.37 (using DeepDB). 
These findings underscore the efficacy of our graph-based representation in providing more accurate cost estimates for UDFs, thereby showcasing the advancements of extensively taking the UDF code structure into account.

\begin{table}
 \centering
 \scalebox{0.88}{
 \begin{tabular}{l|l|rrr}
\textbf{Model} & \textbf{Card. Est. Method} & \multicolumn{1}{c}{\textbf{Median}} & \multicolumn{1}{c}{\textbf{95th}} & \multicolumn{1}{c}{\textbf{99th}} \\ \hline
\system & Actual & 1.29 & 3.58 & 5.17 \\
\system & DeepDB & 1.37 & 7.84 & 25.57 \\ \hline
FlatVector & Actual & 1.89 & 12.66 & 36.10 \\
FlatVector & DeepDB & 2.01 & 17.90 & 344.87 
\end{tabular}}

 \vspace{-1.5ex}
 \caption{Comparing UDF representations on a select-only workload ({\footnotesize \texttt{SELECT udf(col) FROM table WHERE filter}}). The results clearly show the benefits of our approach using a graph-based representation instead of flat representations of UDFs.}
 \label{tab:flat_vector_comparison}
 \vspace{-6ex}
\end{table}

\vspace{-1ex}
\subsection*{Exp 4: Ablation Study}\label{subsec:ablation}
\begin{figure}
 \centering
 \vspace{-2ex}
 \includegraphics[width=.75\linewidth]{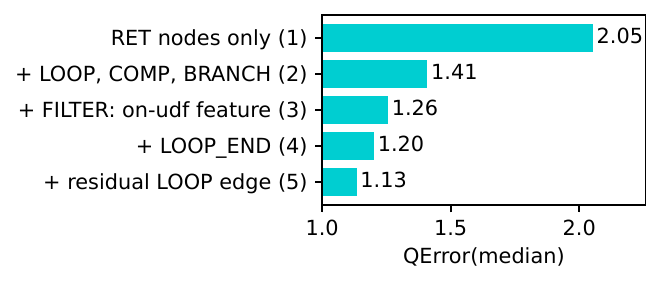}
 \vspace{-2ex}
 \caption{Feature Ablation (actual cardinalities, evaluated on the genome dataset). Comparing different featurizations. (1) UDF represented by only a \texttt{RET} node, (2) add \texttt{LOOP}, \texttt{COMP}, \texttt{BRANCH} and \texttt{INV} nodes, (3) add the on-UDF feature to the \texttt{FILTER}, indicating the filter operation is executed on the output of an UDF, (4) add the \texttt{LOOP\_END} node, (5) add the skip connection between \texttt{LOOP\_START} and \texttt{LOOP\_END} node. 
 }
 \label{fig:feat_ablation}
 \vspace{-4ex}
\end{figure}

\vspace{-.5ex}
To understand how the main design decisions of \system affect cost estimations, we conduct an ablation study as illustrated in \Cref{fig:feat_ablation}. 
For the ablation study, we train multiple models with different sets of features across 19 databases and evaluate their performance on the genome dataset (not included during training). 
Furthermore, we only use actual cardinalities in this experiment to isolate effects from cardinality estimation errors. 

In the ablation study, we first use a simple model (1), keeping only the nodes of the query plan graph and the RET node of the UDF, which indicates only the existence of a UDF. 
This approach models a UDF as a black box as no internal structure is provided. 
In (2), we provide the model with information about the UDF structure by incorporating {\small \texttt{LOOP}, \texttt{COMP}, \texttt{BRANCH}, \texttt{INV}} nodes. 
Further, in (3), we add the UDF-filter feature, which indicates whether the filter operates on the output of a UDF, which provides important information, as discussed before. 
In (4), we then introduce a {\small \texttt{LOOP\_END}} node, which explicitly marks the end of the loop and added a residual edge between the \texttt{LOOP} and {\small \texttt{LOOP\_END}} node in (5) to help the model deal with large graphs.

Since the baseline model (1) contains the query plan without the UDF-specific nodes only, it treats the UDF represented by the {\small \texttt{RET}} node as a black box not having any information about the internal UDF structure. 
Only the number of tuples processed by the UDF is known to the model. 
The resulting median Q-Error for this model is 2.05, highlighting the inherent challenges in estimating runtime without understanding UDF structures and showing how far we can get with accurate estimates of the query plan operators only. 
Next, we incorporate {\small \texttt{LOOP}, \texttt{COMP}, \texttt{BRANCH}} and {\small \texttt{INV}} nodes (2) to provide the model with information about the structure and complexity of the UDF. 
This extension reduces the median Q-Error to 1.41 but increases errors in the higher percentiles, particularly for large UDF graphs. 
This indicates that while structural information improves median performance, it may not fully mitigate challenges with larger UDFs due to the limitations of the \ac{gnn} scheme. 
To address this shortcomings, we introduce a feature for the {\small \texttt{FILTER}} nodes in the query plans that explicitly indicates whether the filter operates on the output of a UDF (3). 
This change reduces the burden on message passing within the model, resulting in a lower median Q-Error of 1.26 and significantly decreasing errors in higher percentiles. 
Furthermore, adding {\small \texttt{LOOP\_END}} nodes (4) to mark the end of loops is especially beneficial for long loops where the information about the loop featurized in the {\small \texttt{LOOP\_START}} node might be diluted by intervening {\small \texttt{COMP}} nodes. 
This adjustment improves the median Q-Error to 1.2 but did not fully resolve issues with large graphs, as the overall graph depth remained challenging for the \ac{gnn}. 
Finally, we introduce residual edges between {\small \texttt{LOOP\_START}} and {\small \texttt{END}} nodes (5) to reduce graph depth, allowing the model to handle large UDFs better. 
This resulted in significant improvements in the higher percentiles: the 95th percentile Q-Error decreased to 4.5, and the 99th to 15.9. 
In summary, these modifications progressively enhance the model's ability to estimate costs accurately by integrating UDF-specific structural and operational features and solving challenges with large graphs introduced with complex and deeply nested UDF code.

\vspace{-.5ex}
\subsection*{Exp 5: Pull-Up Advisor}\label{subsec:exp5_pullup_advisor}
\vspace{-.5ex}

\begin{figure*}
 \centering
 \vspace{-4ex}
 \includegraphics[width=\linewidth]{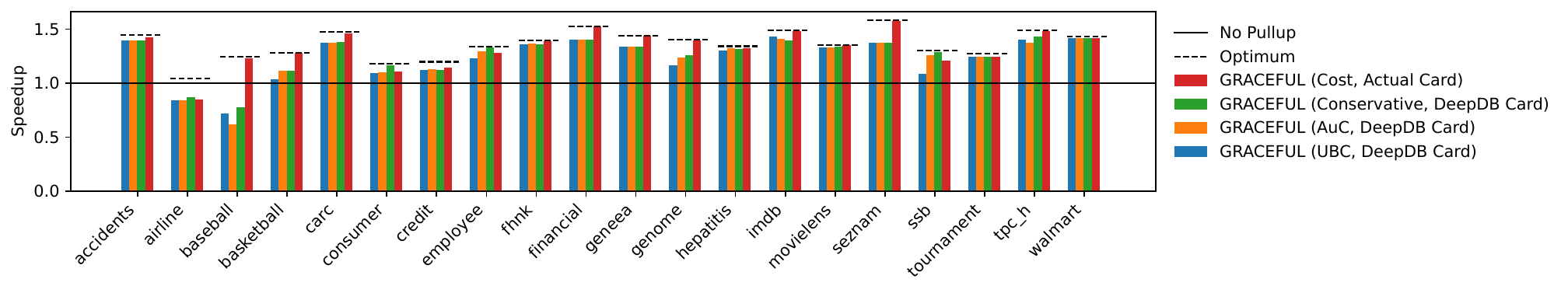}
 \vspace{-5.3ex}
 \caption{Speedups achieved by our Pull-Up advisor with different strategies. Overall, we see robust performance on all datasets and achieve close to optimal speedups. In the usual scenario where no actual cardinalities are available, our conservative advisor strategy shows to overall perform best, reaching similar results although with estimated cardinalities.}
 \label{fig:pull_push_speedups}
 \vspace{-1ex}
\end{figure*}

Finally, we evaluate how \system as a pull-up/pull-down advisor improves the query runtime through two primary experiments designed to asses its performance across diverse datasets and workload scenarios. 
Since our FlatVector baseline supports only select-only workloads, we excluded it from this experiment. 
To test our pull-up advisor we conducted experiments with a subset of 500 queries per dataset where we enforced a complete pull-up to the top and push-down to the bottom (no positioning between) with optimizer hints, while leaving the join order to the query optimizer. 
The experiments are again carried out in a zero-shot setting to measure the advisor's effectiveness on unseen datasets. 

\textbf{Pull-Up Advisor Robustness.}
In the first experiment, we compare the speedups achieved with our proposed advisor with the maximum possible speedup on each of the 20 datasets. 
The experiment is conducted in a leave-one-out cross validation fashion, training each model on 19 datasets and testing it on the unseen 20th. 
By comparing the runtime of the pull-up and push-down cases, we derive the speedups over the workloads.

Our results shown in \Cref{fig:pull_push_speedups} indicate consistent and reliable speedups, underscoring the advisor's capabilities to maintain close to optimal performance despite variations in dataset characteristics. 
Notably, only on the airline and baseball datasets, the advisor's performance face challenges. 
For the baseball dataset, errors in the cardinality annotations lead to less effective decision-making. 
While with actual cards close to optimal performance is reached on this dataset, a significant drop can be seen with estimated cards. 
The reason for this is the lower accuracy of the cardinality estimates on this datasets. 
DeepDB comes with a Q-Error of 1.3 in the median and 30 for the 90th percentile while other datasets show Q-Errors of 1.02 (median) and 8.1 (90th percentile). 
Conversely, the airline dataset exhibits limited potential for speedup, which increased false positives and reduced the advisor's overall efficiency independent of the cardinality estimates or decision strategy in use. 
Despite these outliers, the advisor maintains optimal performance across the majority of datasets, demonstrating its general applicability and resilience. 

\begin{table*}
\begin{center}

 \scalebox{0.88}{
\begin{tabular}{llllllll}
\textbf{Selection Strategy} & \textbf{\begin{tabular}[c]{@{}l@{}}Card. Est. \\ Method\end{tabular}} & \textbf{\begin{tabular}[c]{@{}l@{}}Total Runtime\\ (hrs)\end{tabular}} & \textbf{\begin{tabular}[c]{@{}l@{}}Total \\ Speedup\end{tabular}} & \textbf{\begin{tabular}[c]{@{}l@{}}Median\\ Speedup\end{tabular}} & \textbf{\begin{tabular}[c]{@{}l@{}}False \\ Positives\end{tabular}} & \textbf{\begin{tabular}[c]{@{}l@{}}FP Impact\\ (rel. to total runtime)\end{tabular}} & \textbf{\begin{tabular}[c]{@{}l@{}}Optimization Overhead\\ (rel. to total runtime)\end{tabular}} \\ \hline
Optimal & - & 3.082 & \textbf{1.643} & 1.375 &  - & - & - \\
\system (Cost) & Actual & 3.217 & \textbf{1.574} &1.370 &  0.094 & 0.037 & 3.4 $\%$ \\
\system (Conservative) & DeepDB & 3.460 &  \textbf{1.463} & 1.331 & 0.085 & 0.058 & 3.2 $\%$ \\
\system (AuC) & DeepDB & 3.536 & \textbf{1.432} & 1.329 & 0.117 & 0.079 & 3.1 $\%$ \\
\system (UBC) & DeepDB & 3.595 & \textbf{1.408} & 1.316 & 0.129 & 0.098 & 3.1 $\%$ \\
No Pull-Up (DuckDB Default) & - & 5.063 & 1 & 1 &  0 & 0 & - 
\end{tabular}
}
\vspace{-1.5ex}

 \caption{Speedups of different selection strategies over all 20 datasets. 
 \textit{False Positives} describes the ratio of non-beneficial Pull-Up decisions, while the \textit{FP Impact} describes the impact of these regressions on the total workload (i.e. relative to the total runtime). 
 With actual cardinalities, close to optimal performance can be reached. 
 In the realistic case where the UDF-Filter selectivity can only be estimated, our selection strategies show to achieve large portions of the possible speedup, introducing only little regressions. The runtime of our advisor in comparison to the total query runtime is reasonably low (3.1-3.4\%), while not optimized for online estimation latency.}
 \label{tab:pullup_strategies}
\end{center}
\vspace{-7ex}
\end{table*}

\textbf{Pull-Up Advisor Analysis.}
In the second experiment, we analyze the impact of different pull-up / push-down decision strategies on speedup and regressions across all datasets. 
As shown in \Cref{tab:pullup_strategies}, the speedup with optimal selection of pull-up/push-down is 1.375 compared to the case where no pull-up is applied (which is the usual case for DBMS). 

When applying our model's cost estimates with actual UDF-filter selectivities, the achieved median speedup of 1.370 closely approaches this maximum possible speedup. 
The False Positive Impact (i.e., the slowdown introduced by non-beneficial pull-ups) is minimal, at only 3.7\% relative to the total workload runtime. 
Further, we evaluate the advisor's performance under more realistic conditions where only estimated cardinalities were available. 
Here, we make use of our proposed decision-strategies, which operate on cost distributions instead of a single cost value. 
The conservative strategy achieves a median speedup of 1.33, with a slightly higher regression overhead of 5.8\%, although having no information on the filter selectivity available. 
More greedy strategies, namely AuC and UBC, demonstrate comparable median speedups (1.33 and 1.32) but incur higher regression overheads (8\% and 10\%) eating up the gains of more aggressive pull-ups.
Considering the total runtime of our optimization approach, \system introduces an overhead of only 3.1-3.4\% of the total runtime. 
Since our implementation prioritizes training efficiency over online estimation latency, this overhead could be further reduced. 
Nevertheless, with our optimizations, query runtime decreases from over 5 hours to 3–3.5 hours, depending on the employed strategy. 
This demonstrates that the learned cost model and decision process add only minimal overhead while enabling substantial performance improvements.

\vspace{-.5ex}
\section{Related Work}\label{sec:related_work}
\vspace{-.5ex}

In the following, we discuss related work: First, we focus on optimization strategies for queries with UDF-predicates and afterwards discuss UDF cost estimation.
\vspace{-.75ex}
\subsection{Optimizing Queries with UDF Predicates.}
\vspace{-.5ex}
Addressing the optimization of queries with \acp{udf} predicates has been a recognized challenge since the early 1990s, as initially noted by Stonebraker \cite{DBLP:conf/sigmod/Stonebraker91}. Since then, numerous approaches have aimed to tackle this problem, advancing our understanding of UDFs in query optimization \cite{DBLP:conf/sigmod/Stonebraker91,DBLP:conf/sigmod/HellersteinS93, DBLP:conf/vldb/ChaudhuriS93,DBLP:conf/sigmod/KemperMPS94,DBLP:journals/tods/Hellerstein98,DBLP:conf/vldb/ChaudhuriS96,DBLP:conf/sigmod/JoglekarGPR15}.
Despite these efforts, these approaches generally share two notable limitations: First, they treat UDFs as black boxes, deriving cost information from external sources.
Second, they decouple the cost of the \ac{udf} from the broader query context.

A foundational contribution in this area is the concept of predicate migration by Hellerstein and Stonebraker \cite{DBLP:conf/sigmod/HellersteinS93,DBLP:journals/tods/Hellerstein98}, where they demonstrate that knowledge about the cost of expensive predicates can inform decisions to deviate from traditional predicate push-down. 
Nevertheless, cost estimation in their work is based solely on user-supplied estimates, such as per-invocation execution time. 
Similarly, Chaudhuri and Shim's approach \cite{DBLP:conf/vldb/ChaudhuriS96} models UDF complexity using user-provided data, such as expected I/O access patterns. 
In another related direction, Kemper et al. \cite{DBLP:conf/sigmod/KemperMPS94} address the problem of predicate reordering, prioritizing the evaluation of less costly predicates. 
However, they also assume the costs of these expensive predicates as known beforehand.

\vspace{-.75ex}
\subsection{UDF Cost Estimation}
\vspace{-.5ex}

Approaches to UDF cost estimation, in general, assume that the \ac{udf} has been seen before and runtime statistics are available or that CPU costs are manually annotated. 
This is in stark contrast to our work, which focuses on generating predictions for unseen datasets. 
The existing approaches can be grouped into three primary categories: 

\textbf{Statistical Approaches for UDF Cost Estimation.}
Early approaches, such as \cite{DBLP:journals/sigmod/BoulosO99,boulos1997neural}, treated UDF logic as black boxes. 
These models leverage curve-fitting techniques and neural networks to predict costs by training on execution times across various input parameters.
\cite{DBLP:journals/sigmod/BoulosO99} advances this with multi-dimensional histograms, capturing execution times across varying input configurations and storing them in a hierarchal tree structure for prediction. 

\textbf{Self-Tuning Models for UDF Cost Estimation.}
Self-tuning models, such as \cite{DBLP:journals/cj/LeeCBK04,DBLP:journals/tods/HeLS05}, adjust dynamically based on observed UDF executions, updating cost estimates with each new UDF run.
For this, \cite{DBLP:journals/cj/LeeCBK04} proposed a regression-based model that refines cost estimates with each UDF execution.
Building on this, \cite{DBLP:journals/tods/HeLS05} introduces KNN and memory-limited quadtrees (MLQ) techniques, which improve cost estimation by aligning historical UDF input values with observed costs. 

\textbf{Static Code Analysis.}
\cite{DBLP:conf/icde/HueskePKRTMF13} uses static analysis to optimize Map-Reduce-style UDFs in data flow programs. 
By analyzing control and data flow graphs, they extract read and write sets and predict output cardinalities for efficient operator reordering. 
To achieve accurate cost estimates, CPU costs and selecitivities are manually annotated or derived by profiling.

\textbf{Representing Code.}
Representing program code as graphs is a well-established approach in programming languages and machine learning. 

Key methods include abstract syntax trees \cite{DBLP:journals/pacmpl/AlonZLY19,DBLP:conf/iclr/AlonBLY19,DBLP:conf/icml/0002SLY20,DBLP:conf/icse/KimZT021,DBLP:conf/icse/Li0N21a,DBLP:conf/aaai/WangL21a}, control flow graphs \cite{DBLP:conf/iclr/BieberGZLT23,DBLP:conf/sigsoft/DeFreezTR18,DBLP:conf/ictai/PhanNB17,DBLP:journals/virology/AndersonQNSL11, DBLP:conf/dimva/BruschiMM06, DBLP:conf/cikm/ChaeHKKI13, DBLP:conf/sec/SunZXMX14} and data flow graphs \cite{DBLP:conf/iclr/GuoRLFT0ZDSFTDC21,DBLP:conf/mlsys/KaufmanPZM0SB21}.
This diverse body of work underlines the complexity of UDF cost estimation and underscores the variety of techniques developed to optimize \acp{udf}, reflecting the evolving needs of query optimization in modern \ac{dbms}.
\vspace{-.5ex}
\section{Conclusion and Future Work}\label{sec:conclusion}
\vspace{-.5ex}
Our work addresses a critical gap in current database query optimization, where support for UDFs is often limited or non-existent. 
In this paper, we have demonstrated a novel approach for accurately and robustly predicting the runtime costs of User-Defined Functions (UDFs) in SQL queries. Additionally, we proved the usability of our cost estimator for real-world scenarios by applying it to the task of pull-up vs. push-down optimizations. 
For this, we build a pull-up advisor on top of our cost estimator, which predicts optimal strategies for UDF pull-up and push-down within SQL queries, consistently achieving significant speedups across 20 diverse datasets with minimal performance regressions. 
These experimental evaluations confirm the effectiveness of our method across various complexities and dataset conditions and its usability for real-world tasks. 
While we believe that our work is an important step to enable query optimization for queries with UDFs, more work is needed in particular on how to enable cost-based optimizations for UDFs that go beyond pull-up/push-down decisions.
For this, we published our code and the benchmark to support further research in this area.

\bibliographystyle{IEEEtran}
\balance
\bibliography{main.bib}

\begin{thebibliography}{10}
\providecommand{\url}[1]{#1}
\csname url@samestyle\endcsname
\providecommand{\newblock}{\relax}
\providecommand{\bibinfo}[2]{#2}
\providecommand{\BIBentrySTDinterwordspacing}{\spaceskip=0pt\relax}
\providecommand{\BIBentryALTinterwordstretchfactor}{4}
\providecommand{\BIBentryALTinterwordspacing}{\spaceskip=\fontdimen2\font plus
\BIBentryALTinterwordstretchfactor\fontdimen3\font minus
  \fontdimen4\font\relax}
\providecommand{\BIBforeignlanguage}[2]{{%
\expandafter\ifx\csname l@#1\endcsname\relax
\typeout{** WARNING: IEEEtran.bst: No hyphenation pattern has been}%
\typeout{** loaded for the language `#1'. Using the pattern for}%
\typeout{** the default language instead.}%
\else
\language=\csname l@#1\endcsname
\fi
#2}}
\providecommand{\BIBdecl}{\relax}
\BIBdecl

\bibitem{DBLP:journals/pvldb/GuptaR21}
\BIBentryALTinterwordspacing
S.~Gupta and K.~Ramachandra, ``Procedural extensions of {SQL:} understanding
  their usage in the wild,'' \emph{Proc. {VLDB} Endow.}, vol.~14, no.~8, pp.
  1378--1391, 2021. [Online]. Available:
  \url{http://www.vldb.org/pvldb/vol14/p1378-ramachandra.pdf}
\BIBentrySTDinterwordspacing

\bibitem{DBLP:journals/pvldb/RamachandraPEHG17}
\BIBentryALTinterwordspacing
K.~Ramachandra, K.~Park, K.~V. Emani, A.~Halverson, C.~A. Galindo{-}Legaria,
  and C.~Cunningham, ``Froid: Optimization of imperative programs in a
  relational database,'' \emph{Proc. {VLDB} Endow.}, vol.~11, no.~4, pp.
  432--444, 2017. [Online]. Available:
  \url{http://www.vldb.org/pvldb/vol11/p432-ramachandra.pdf}
\BIBentrySTDinterwordspacing

\bibitem{DBLP:conf/sigmod/Stonebraker91}
\BIBentryALTinterwordspacing
M.~Stonebraker, ``Managing persistent objects in a multi-level store,'' in
  \emph{Proceedings of the 1991 {ACM} {SIGMOD} International Conference on
  Management of Data, Denver, Colorado, USA, May 29-31, 1991}, J.~Clifford and
  R.~King, Eds.\hskip 1em plus 0.5em minus 0.4em\relax {ACM} Press, 1991, pp.
  2--11. [Online]. Available: \url{https://doi.org/10.1145/115790.115791}
\BIBentrySTDinterwordspacing

\bibitem{DBLP:conf/cidr/FranzAHGMP24}
\BIBentryALTinterwordspacing
K.~Franz, S.~Arch, D.~Hirn, T.~Grust, T.~C. Mowry, and A.~Pavlo, ``Dear
  user-defined functions, inlining isn't working out so great for us. let's try
  batching to make our relationship work. sincerely, {SQL},'' in \emph{14th
  Conference on Innovative Data Systems Research, {CIDR} 2024, Chaminade, HI,
  USA, January 14-17, 2024}.\hskip 1em plus 0.5em minus 0.4em\relax
  www.cidrdb.org, 2024. [Online]. Available:
  \url{https://www.cidrdb.org/cidr2024/papers/p13-franz.pdf}
\BIBentrySTDinterwordspacing

\bibitem{DBLP:journals/pvldb/LeisGMBK015}
\BIBentryALTinterwordspacing
V.~Leis, A.~Gubichev, A.~Mirchev, P.~A. Boncz, A.~Kemper, and T.~Neumann, ``How
  good are query optimizers, really?'' \emph{Proc. {VLDB} Endow.}, vol.~9,
  no.~3, pp. 204--215, 2015. [Online]. Available:
  \url{http://www.vldb.org/pvldb/vol9/p204-leis.pdf}
\BIBentrySTDinterwordspacing

\bibitem{DBLP:conf/sigmod/ZhangIM0GLFHPJ22}
\BIBentryALTinterwordspacing
W.~Zhang, M.~Interlandi, P.~Mineiro, S.~Qiao, N.~Ghazanfari, K.~Lie, M.~T.
  Friedman, R.~Hosn, H.~Patel, and A.~Jindal, ``Deploying a steered query
  optimizer in production at microsoft,'' in \emph{{SIGMOD} '22: International
  Conference on Management of Data, Philadelphia, PA, USA, June 12 - 17, 2022},
  Z.~G. Ives, A.~Bonifati, and A.~E. Abbadi, Eds.\hskip 1em plus 0.5em minus
  0.4em\relax {ACM}, 2022, pp. 2299--2311. [Online]. Available:
  \url{https://doi.org/10.1145/3514221.3526052}
\BIBentrySTDinterwordspacing

\bibitem{DBLP:books/mg/Ramakrishnan98}
\BIBentryALTinterwordspacing
R.~Ramakrishnan, \emph{Database Management Systems}.\hskip 1em plus 0.5em minus
  0.4em\relax WCB/McGraw-Hill, 1998. [Online]. Available:
  \url{http://www.cs.wisc.edu/\%7Edbbook}
\BIBentrySTDinterwordspacing

\bibitem{DBLP:books/daglib/0010423}
H.~Garcia{-}Molina, J.~D. Ullman, and J.~Widom, \emph{Database systems - the
  complete book (international edition)}.\hskip 1em plus 0.5em minus
  0.4em\relax Pearson Education, 2002.

\bibitem{DBLP:conf/icde/LiangCXYCX024}
\BIBentryALTinterwordspacing
Z.~Liang, X.~Chen, Y.~Xia, R.~Ye, H.~Chen, J.~Xie, and K.~Zheng, ``{DACE:} {A}
  database-agnostic cost estimator,'' in \emph{40th {IEEE} International
  Conference on Data Engineering, {ICDE} 2024, Utrecht, The Netherlands, May
  13-16, 2024}.\hskip 1em plus 0.5em minus 0.4em\relax {IEEE}, 2024, pp.
  4925--4937. [Online]. Available:
  \url{https://doi.org/10.1109/ICDE60146.2024.00374}
\BIBentrySTDinterwordspacing

\bibitem{DBLP:conf/sigmod/WuMLNNPSRNK24}
\BIBentryALTinterwordspacing
Z.~Wu, R.~Marcus, Z.~Liu, P.~Negi, V.~Nathan, P.~Pfeil, G.~Saxena, M.~Rahman,
  B.~Narayanaswamy, and T.~Kraska, ``Stage: Query execution time prediction in
  amazon redshift,'' in \emph{Companion of the 2024 International Conference on
  Management of Data, {SIGMOD/PODS} 2024, Santiago AA, Chile, June 9-15, 2024},
  P.~Barcel{\'{o}}, N.~S{\'{a}}nchez{-}Pi, A.~Meliou, and S.~Sudarshan,
  Eds.\hskip 1em plus 0.5em minus 0.4em\relax {ACM}, 2024, pp. 280--294.
  [Online]. Available: \url{https://doi.org/10.1145/3626246.3653391}
\BIBentrySTDinterwordspacing

\bibitem{DBLP:journals/pvldb/HilprechtB22}
\BIBentryALTinterwordspacing
B.~Hilprecht and C.~Binnig, ``Zero-shot cost models for out-of-the-box learned
  cost prediction,'' \emph{Proc. {VLDB} Endow.}, vol.~15, no.~11, pp.
  2361--2374, 2022. [Online]. Available:
  \url{https://www.vldb.org/pvldb/vol15/p2361-hilprecht.pdf}
\BIBentrySTDinterwordspacing

\bibitem{DBLP:journals/pvldb/ZhaoCSM22}
\BIBentryALTinterwordspacing
Y.~Zhao, G.~Cong, J.~Shi, and C.~Miao, ``Queryformer: {A} tree transformer
  model for query plan representation,'' \emph{Proc. {VLDB} Endow.}, vol.~15,
  no.~8, pp. 1658--1670, 2022. [Online]. Available:
  \url{https://www.vldb.org/pvldb/vol15/p1658-zhao.pdf}
\BIBentrySTDinterwordspacing

\bibitem{DBLP:journals/pvldb/SunL19}
\BIBentryALTinterwordspacing
J.~Sun and G.~Li, ``An end-to-end learning-based cost estimator,'' \emph{Proc.
  {VLDB} Endow.}, vol.~13, no.~3, pp. 307--319, 2019. [Online]. Available:
  \url{http://www.vldb.org/pvldb/vol13/p307-sun.pdf}
\BIBentrySTDinterwordspacing

\bibitem{DBLP:journals/pvldb/MarcusP19}
\BIBentryALTinterwordspacing
R.~Marcus and O.~Papaemmanouil, ``Plan-structured deep neural network models
  for query performance prediction,'' \emph{Proc. {VLDB} Endow.}, vol.~12,
  no.~11, pp. 1733--1746, 2019. [Online]. Available:
  \url{http://www.vldb.org/pvldb/vol12/p1733-marcus.pdf}
\BIBentrySTDinterwordspacing

\bibitem{DBLP:conf/icde/HeinrichBKL24}
\BIBentryALTinterwordspacing
R.~Heinrich, C.~Binnig, H.~Kornmayer, and M.~Luthra, ``Costream: Learned cost
  models for operator placement in edge-cloud environments,'' in \emph{40th
  {IEEE} International Conference on Data Engineering, {ICDE} 2024, Utrecht,
  The Netherlands, May 13-16, 2024}.\hskip 1em plus 0.5em minus 0.4em\relax
  {IEEE}, 2024, pp. 96--109. [Online]. Available:
  \url{https://doi.org/10.1109/ICDE60146.2024.00015}
\BIBentrySTDinterwordspacing

\bibitem{DBLP:journals/pvldb/MarcusNMZAKPT19}
\BIBentryALTinterwordspacing
R.~Marcus, P.~Negi, H.~Mao, C.~Zhang, M.~Alizadeh, T.~Kraska, O.~Papaemmanouil,
  and N.~Tatbul, ``Neo: {A} learned query optimizer,'' \emph{Proc. {VLDB}
  Endow.}, vol.~12, no.~11, pp. 1705--1718, 2019. [Online]. Available:
  \url{http://www.vldb.org/pvldb/vol12/p1705-marcus.pdf}
\BIBentrySTDinterwordspacing

\bibitem{DBLP:journals/sigmod/MarcusNMTAK22}
\BIBentryALTinterwordspacing
R.~Marcus, P.~Negi, H.~Mao, N.~Tatbul, M.~Alizadeh, and T.~Kraska, ``Bao:
  Making learned query optimization practical,'' \emph{{SIGMOD} Rec.}, vol.~51,
  no.~1, pp. 6--13, 2022. [Online]. Available:
  \url{https://doi.org/10.1145/3542700.3542703}
\BIBentrySTDinterwordspacing

\bibitem{DBLP:journals/pvldb/ZhuCDCPWZ23}
\BIBentryALTinterwordspacing
R.~Zhu, W.~Chen, B.~Ding, X.~Chen, A.~Pfadler, Z.~Wu, and J.~Zhou, ``Lero: {A}
  learning-to-rank query optimizer,'' \emph{Proc. {VLDB} Endow.}, vol.~16,
  no.~6, pp. 1466--1479, 2023. [Online]. Available:
  \url{https://www.vldb.org/pvldb/vol16/p1466-zhu.pdf}
\BIBentrySTDinterwordspacing

\bibitem{DBLP:journals/pacmmod/DoshiZJMHABF23}
\BIBentryALTinterwordspacing
L.~Doshi, V.~Zhuang, G.~Jain, R.~Marcus, H.~Huang, D.~Altinb{\"{u}}ken,
  E.~Brevdo, and C.~Fraser, ``Kepler: Robust learning for parametric query
  optimization,'' \emph{Proc. {ACM} Manag. Data}, vol.~1, no.~1, pp.
  109:1--109:25, 2023. [Online]. Available:
  \url{https://doi.org/10.1145/3588963}
\BIBentrySTDinterwordspacing

\bibitem{DBLP:conf/cidr/HilprechtB22}
\BIBentryALTinterwordspacing
B.~Hilprecht and C.~Binnig, ``One model to rule them all: Towards zero-shot
  learning for databases,'' in \emph{12th Conference on Innovative Data Systems
  Research, {CIDR} 2022, Chaminade, CA, USA, January 9-12, 2022}.\hskip 1em
  plus 0.5em minus 0.4em\relax www.cidrdb.org, 2022. [Online]. Available:
  \url{https://www.cidrdb.org/cidr2022/papers/p16-hilprecht.pdf}
\BIBentrySTDinterwordspacing

\bibitem{DBLP:journals/corr/abs-2208-07461}
\BIBentryALTinterwordspacing
D.~Bieber, K.~Shi, P.~Maniatis, C.~Sutton, V.~J. Hellendoorn, D.~D. Johnson,
  and D.~Tarlow, ``A library for representing python programs as graphs for
  machine learning,'' \emph{CoRR}, vol. abs/2208.07461, 2022. [Online].
  Available: \url{https://doi.org/10.48550/arXiv.2208.07461}
\BIBentrySTDinterwordspacing

\bibitem{DBLP:journals/pvldb/LeeDNC23}
\BIBentryALTinterwordspacing
K.~Lee, A.~Dutt, V.~R. Narasayya, and S.~Chaudhuri, ``Analyzing the impact of
  cardinality estimation on execution plans in microsoft {SQL} server,''
  \emph{Proc. {VLDB} Endow.}, vol.~16, no.~11, pp. 2871--2883, 2023. [Online].
  Available: \url{https://www.vldb.org/pvldb/vol16/p2871-dutt.pdf}
\BIBentrySTDinterwordspacing

\bibitem{DBLP:conf/cidr/KipfKRLBK19}
\BIBentryALTinterwordspacing
A.~Kipf, T.~Kipf, B.~Radke, V.~Leis, P.~A. Boncz, and A.~Kemper, ``Learned
  cardinalities: Estimating correlated joins with deep learning,'' in \emph{9th
  Biennial Conference on Innovative Data Systems Research, {CIDR} 2019,
  Asilomar, CA, USA, January 13-16, 2019, Online Proceedings}.\hskip 1em plus
  0.5em minus 0.4em\relax www.cidrdb.org, 2019. [Online]. Available:
  \url{http://cidrdb.org/cidr2019/papers/p101-kipf-cidr19.pdf}
\BIBentrySTDinterwordspacing

\bibitem{DBLP:journals/pvldb/RenenHPVDNLSKK24}
\BIBentryALTinterwordspacing
A.~van Renen, D.~Horn, P.~Pfeil, K.~Vaidya, W.~Dong, M.~Narayanaswamy, Z.~Liu,
  G.~Saxena, A.~Kipf, and T.~Kraska, ``Why {TPC} is not enough: An analysis of
  the amazon redshift fleet,'' \emph{Proc. {VLDB} Endow.}, vol.~17, no.~11, pp.
  3694--3706, 2024. [Online]. Available:
  \url{https://www.vldb.org/pvldb/vol17/p3694-saxena.pdf}
\BIBentrySTDinterwordspacing

\bibitem{DBLP:conf/sigmod/RaasveldtM19}
\BIBentryALTinterwordspacing
M.~Raasveldt and H.~M{\"{u}}hleisen, ``Duckdb: an embeddable analytical
  database,'' in \emph{Proceedings of the 2019 International Conference on
  Management of Data, {SIGMOD} Conference 2019, Amsterdam, The Netherlands,
  June 30 - July 5, 2019}, P.~A. Boncz, S.~Manegold, A.~Ailamaki, A.~Deshpande,
  and T.~Kraska, Eds.\hskip 1em plus 0.5em minus 0.4em\relax {ACM}, 2019, pp.
  1981--1984. [Online]. Available:
  \url{https://doi.org/10.1145/3299869.3320212}
\BIBentrySTDinterwordspacing

\bibitem{DBLP:journals/pvldb/HilprechtSKMKB20}
\BIBentryALTinterwordspacing
B.~Hilprecht, A.~Schmidt, M.~Kulessa, A.~Molina, K.~Kersting, and C.~Binnig,
  ``Deepdb: Learn from data, not from queries!'' \emph{Proc. {VLDB} Endow.},
  vol.~13, no.~7, pp. 992--1005, 2020. [Online]. Available:
  \url{http://www.vldb.org/pvldb/vol13/p992-hilprecht.pdf}
\BIBentrySTDinterwordspacing

\bibitem{DBLP:conf/sigmod/0001WYZ16}
\BIBentryALTinterwordspacing
F.~Li, B.~Wu, K.~Yi, and Z.~Zhao, ``Wander join: Online aggregation via random
  walks,'' in \emph{Proceedings of the 2016 International Conference on
  Management of Data, {SIGMOD} Conference 2016, San Francisco, CA, USA, June 26
  - July 01, 2016}, F.~{\"{O}}zcan, G.~Koutrika, and S.~Madden, Eds.\hskip 1em
  plus 0.5em minus 0.4em\relax {ACM}, 2016, pp. 615--629. [Online]. Available:
  \url{https://doi.org/10.1145/2882903.2915235}
\BIBentrySTDinterwordspacing

\bibitem{boulos1997neural}
J.~Boulos, Y.~Viemont, and K.~Ono, ``A neural networks approach for query cost
  evaluation,'' \emph{Transaction of Information Processing Society of Japan},
  vol.~38, no.~12, pp. 2566--2575, 1997.

\bibitem{DBLP:conf/icde/GanapathiKDWFJP09}
\BIBentryALTinterwordspacing
A.~Ganapathi, H.~A. Kuno, U.~Dayal, J.~L. Wiener, A.~Fox, M.~I. Jordan, and
  D.~A. Patterson, ``Predicting multiple metrics for queries: Better decisions
  enabled by machine learning,'' in \emph{Proceedings of the 25th International
  Conference on Data Engineering, {ICDE} 2009, March 29 2009 - April 2 2009,
  Shanghai, China}, Y.~E. Ioannidis, D.~L. Lee, and R.~T. Ng, Eds.\hskip 1em
  plus 0.5em minus 0.4em\relax {IEEE} Computer Society, 2009, pp. 592--603.
  [Online]. Available: \url{https://doi.org/10.1109/ICDE.2009.130}
\BIBentrySTDinterwordspacing

\bibitem{heinrich2025goodlearnedcostmodels}
\BIBentryALTinterwordspacing
R.~Heinrich, M.~Luthra, J.~Wehrstein, H.~Kornmayer, and C.~Binnig, ``How good
  are learned cost models, really? insights from query optimization tasks,''
  2025. [Online]. Available: \url{https://arxiv.org/abs/2502.01229}
\BIBentrySTDinterwordspacing

\bibitem{DBLP:conf/kdd/ChenG16}
\BIBentryALTinterwordspacing
T.~Chen and C.~Guestrin, ``Xgboost: {A} scalable tree boosting system,'' in
  \emph{Proceedings of the 22nd {ACM} {SIGKDD} International Conference on
  Knowledge Discovery and Data Mining, San Francisco, CA, USA, August 13-17,
  2016}, B.~Krishnapuram, M.~Shah, A.~J. Smola, C.~C. Aggarwal, D.~Shen, and
  R.~Rastogi, Eds.\hskip 1em plus 0.5em minus 0.4em\relax {ACM}, 2016, pp.
  785--794. [Online]. Available: \url{https://doi.org/10.1145/2939672.2939785}
\BIBentrySTDinterwordspacing

\bibitem{DBLP:conf/sigmod/HellersteinS93}
\BIBentryALTinterwordspacing
J.~M. Hellerstein and M.~Stonebraker, ``Predicate migration: Optimizing queries
  with expensive predicates,'' in \emph{Proceedings of the 1993 {ACM} {SIGMOD}
  International Conference on Management of Data, Washington, DC, USA, May
  26-28, 1993}, P.~Buneman and S.~Jajodia, Eds.\hskip 1em plus 0.5em minus
  0.4em\relax {ACM} Press, 1993, pp. 267--276. [Online]. Available:
  \url{https://doi.org/10.1145/170035.170078}
\BIBentrySTDinterwordspacing

\bibitem{DBLP:conf/vldb/ChaudhuriS93}
\BIBentryALTinterwordspacing
S.~Chaudhuri and K.~Shim, ``Query optimization in the presence of foreign
  functions,'' in \emph{19th International Conference on Very Large Data Bases,
  August 24-27, 1993, Dublin, Ireland, Proceedings}, R.~Agrawal, S.~Baker, and
  D.~A. Bell, Eds.\hskip 1em plus 0.5em minus 0.4em\relax Morgan Kaufmann,
  1993, pp. 529--542. [Online]. Available:
  \url{http://www.vldb.org/conf/1993/P529.PDF}
\BIBentrySTDinterwordspacing

\bibitem{DBLP:conf/sigmod/KemperMPS94}
\BIBentryALTinterwordspacing
A.~Kemper, G.~Moerkotte, K.~Peithner, and M.~Steinbrunn, ``Optimizing
  disjunctive queries with expensive predicates,'' in \emph{Proceedings of the
  1994 {ACM} {SIGMOD} International Conference on Management of Data,
  Minneapolis, Minnesota, USA, May 24-27, 1994}, R.~T. Snodgrass and
  M.~Winslett, Eds.\hskip 1em plus 0.5em minus 0.4em\relax {ACM} Press, 1994,
  pp. 336--347. [Online]. Available:
  \url{https://doi.org/10.1145/191839.191906}
\BIBentrySTDinterwordspacing

\bibitem{DBLP:journals/tods/Hellerstein98}
\BIBentryALTinterwordspacing
J.~M. Hellerstein, ``Optimization techniques for queries with expensive
  methods,'' \emph{{ACM} Trans. Database Syst.}, vol.~23, no.~2, pp. 113--157,
  1998. [Online]. Available: \url{https://doi.org/10.1145/292481.277627}
\BIBentrySTDinterwordspacing

\bibitem{DBLP:conf/vldb/ChaudhuriS96}
\BIBentryALTinterwordspacing
S.~Chaudhuri and K.~Shim, ``Optimization of queries with user-defined
  predicates,'' in \emph{VLDB'96, Proceedings of 22th International Conference
  on Very Large Data Bases, September 3-6, 1996, Mumbai (Bombay), India}, T.~M.
  Vijayaraman, A.~P. Buchmann, C.~Mohan, and N.~L. Sarda, Eds.\hskip 1em plus
  0.5em minus 0.4em\relax Morgan Kaufmann, 1996, pp. 87--98. [Online].
  Available: \url{http://www.vldb.org/conf/1996/P087.PDF}
\BIBentrySTDinterwordspacing

\bibitem{DBLP:conf/sigmod/JoglekarGPR15}
\BIBentryALTinterwordspacing
M.~Joglekar, H.~Garcia{-}Molina, A.~G. Parameswaran, and C.~R{\'{e}},
  ``Exploiting correlations for expensive predicate evaluation,'' in
  \emph{Proceedings of the 2015 {ACM} {SIGMOD} International Conference on
  Management of Data, Melbourne, Victoria, Australia, May 31 - June 4, 2015},
  T.~K. Sellis, S.~B. Davidson, and Z.~G. Ives, Eds.\hskip 1em plus 0.5em minus
  0.4em\relax {ACM}, 2015, pp. 1183--1198. [Online]. Available:
  \url{https://doi.org/10.1145/2723372.2723715}
\BIBentrySTDinterwordspacing

\bibitem{DBLP:journals/sigmod/BoulosO99}
\BIBentryALTinterwordspacing
J.~Boulos and K.~Ono, ``Cost estimation of user-defined methods in
  object-relational database systems,'' \emph{{SIGMOD} Rec.}, vol.~28, no.~3,
  pp. 22--28, 1999. [Online]. Available:
  \url{https://doi.org/10.1145/333607.333610}
\BIBentrySTDinterwordspacing

\bibitem{DBLP:journals/cj/LeeCBK04}
\BIBentryALTinterwordspacing
B.~S. Lee, L.~Chen, J.~Buzas, and V.~Kannoth, ``Regression-based self-tuning
  modeling of smooth user-defined function costs for an object-relational
  database management system query optimizer,'' \emph{Comput. J.}, vol.~47,
  no.~6, pp. 673--693, 2004. [Online]. Available:
  \url{https://doi.org/10.1093/comjnl/47.6.673}
\BIBentrySTDinterwordspacing

\bibitem{DBLP:journals/tods/HeLS05}
\BIBentryALTinterwordspacing
Z.~He, B.~S. Lee, and R.~R. Snapp, ``Self-tuning cost modeling of user-defined
  functions in an object-relational {DBMS},'' \emph{{ACM} Trans. Database
  Syst.}, vol.~30, no.~3, pp. 812--853, 2005. [Online]. Available:
  \url{https://doi.org/10.1145/1093382.1093387}
\BIBentrySTDinterwordspacing

\bibitem{DBLP:conf/icde/HueskePKRTMF13}
\BIBentryALTinterwordspacing
F.~Hueske, M.~Peters, A.~Krettek, M.~Ringwald, K.~Tzoumas, V.~Markl, and
  J.~Freytag, ``Peeking into the optimization of data flow programs with
  mapreduce-style udfs,'' in \emph{29th {IEEE} International Conference on Data
  Engineering, {ICDE} 2013, Brisbane, Australia, April 8-12, 2013}, C.~S.
  Jensen, C.~M. Jermaine, and X.~Zhou, Eds.\hskip 1em plus 0.5em minus
  0.4em\relax {IEEE} Computer Society, 2013, pp. 1292--1295. [Online].
  Available: \url{https://doi.org/10.1109/ICDE.2013.6544927}
\BIBentrySTDinterwordspacing

\bibitem{DBLP:journals/pacmpl/AlonZLY19}
\BIBentryALTinterwordspacing
U.~Alon, M.~Zilberstein, O.~Levy, and E.~Yahav, ``code2vec: learning
  distributed representations of code,'' \emph{Proc. {ACM} Program. Lang.},
  vol.~3, no. {POPL}, pp. 40:1--40:29, 2019. [Online]. Available:
  \url{https://doi.org/10.1145/3290353}
\BIBentrySTDinterwordspacing

\bibitem{DBLP:conf/iclr/AlonBLY19}
\BIBentryALTinterwordspacing
U.~Alon, S.~Brody, O.~Levy, and E.~Yahav, ``code2seq: Generating sequences from
  structured representations of code,'' in \emph{7th International Conference
  on Learning Representations, {ICLR} 2019, New Orleans, LA, USA, May 6-9,
  2019}.\hskip 1em plus 0.5em minus 0.4em\relax OpenReview.net, 2019. [Online].
  Available: \url{https://openreview.net/forum?id=H1gKYo09tX}
\BIBentrySTDinterwordspacing

\bibitem{DBLP:conf/icml/0002SLY20}
\BIBentryALTinterwordspacing
U.~Alon, R.~Sadaka, O.~Levy, and E.~Yahav, ``Structural language models of
  code,'' in \emph{Proceedings of the 37th International Conference on Machine
  Learning, {ICML} 2020, 13-18 July 2020, Virtual Event}, ser. Proceedings of
  Machine Learning Research, vol. 119.\hskip 1em plus 0.5em minus 0.4em\relax
  {PMLR}, 2020, pp. 245--256. [Online]. Available:
  \url{http://proceedings.mlr.press/v119/alon20a.html}
\BIBentrySTDinterwordspacing

\bibitem{DBLP:conf/icse/KimZT021}
\BIBentryALTinterwordspacing
S.~Kim, J.~Zhao, Y.~Tian, and S.~Chandra, ``Code prediction by feeding trees to
  transformers,'' in \emph{43rd {IEEE/ACM} International Conference on Software
  Engineering, {ICSE} 2021, Madrid, Spain, 22-30 May 2021}.\hskip 1em plus
  0.5em minus 0.4em\relax {IEEE}, 2021, pp. 150--162. [Online]. Available:
  \url{https://doi.org/10.1109/ICSE43902.2021.00026}
\BIBentrySTDinterwordspacing

\bibitem{DBLP:conf/icse/Li0N21a}
\BIBentryALTinterwordspacing
Y.~Li, S.~Wang, and T.~N. Nguyen, ``Fault localization with code coverage
  representation learning,'' in \emph{43rd {IEEE/ACM} International Conference
  on Software Engineering, {ICSE} 2021, Madrid, Spain, 22-30 May 2021}.\hskip
  1em plus 0.5em minus 0.4em\relax {IEEE}, 2021, pp. 661--673. [Online].
  Available: \url{https://doi.org/10.1109/ICSE43902.2021.00067}
\BIBentrySTDinterwordspacing

\bibitem{DBLP:conf/aaai/WangL21a}
\BIBentryALTinterwordspacing
Y.~Wang and H.~Li, ``Code completion by modeling flattened abstract syntax
  trees as graphs,'' in \emph{Thirty-Fifth {AAAI} Conference on Artificial
  Intelligence, {AAAI} 2021, Thirty-Third Conference on Innovative Applications
  of Artificial Intelligence, {IAAI} 2021, The Eleventh Symposium on
  Educational Advances in Artificial Intelligence, {EAAI} 2021, Virtual Event,
  February 2-9, 2021}.\hskip 1em plus 0.5em minus 0.4em\relax {AAAI} Press,
  2021, pp. 14\,015--14\,023. [Online]. Available:
  \url{https://doi.org/10.1609/aaai.v35i16.17650}
\BIBentrySTDinterwordspacing

\bibitem{DBLP:conf/iclr/BieberGZLT23}
\BIBentryALTinterwordspacing
D.~Bieber, R.~Goel, D.~Zheng, H.~Larochelle, and D.~Tarlow, ``Static prediction
  of runtime errors by learning to execute programs with external resource
  descriptions,'' in \emph{The Eleventh International Conference on Learning
  Representations, {ICLR} 2023, Kigali, Rwanda, May 1-5, 2023}.\hskip 1em plus
  0.5em minus 0.4em\relax OpenReview.net, 2023. [Online]. Available:
  \url{https://openreview.net/pdf?id=lLp-C5nTdJG}
\BIBentrySTDinterwordspacing

\bibitem{DBLP:conf/sigsoft/DeFreezTR18}
\BIBentryALTinterwordspacing
D.~DeFreez, A.~V. Thakur, and C.~Rubio{-}Gonz{\'{a}}lez, ``Path-based function
  embedding and its application to error-handling specification mining,'' in
  \emph{Proceedings of the 2018 {ACM} Joint Meeting on European Software
  Engineering Conference and Symposium on the Foundations of Software
  Engineering, {ESEC/SIGSOFT} {FSE} 2018, Lake Buena Vista, FL, USA, November
  04-09, 2018}, G.~T. Leavens, A.~Garcia, and C.~S. Pasareanu, Eds.\hskip 1em
  plus 0.5em minus 0.4em\relax {ACM}, 2018, pp. 423--433. [Online]. Available:
  \url{https://doi.org/10.1145/3236024.3236059}
\BIBentrySTDinterwordspacing

\bibitem{DBLP:conf/ictai/PhanNB17}
\BIBentryALTinterwordspacing
A.~V. Phan, M.~L. Nguyen, and L.~T. Bui, ``Convolutional neural networks over
  control flow graphs for software defect prediction,'' in \emph{29th {IEEE}
  International Conference on Tools with Artificial Intelligence, {ICTAI} 2017,
  Boston, MA, USA, November 6-8, 2017}.\hskip 1em plus 0.5em minus 0.4em\relax
  {IEEE} Computer Society, 2017, pp. 45--52. [Online]. Available:
  \url{https://doi.org/10.1109/ICTAI.2017.00019}
\BIBentrySTDinterwordspacing

\bibitem{DBLP:journals/virology/AndersonQNSL11}
\BIBentryALTinterwordspacing
B.~Anderson, D.~Quist, J.~Neil, C.~B. Storlie, and T.~Lane, ``Graph-based
  malware detection using dynamic analysis,'' \emph{J. Comput. Virol.}, vol.~7,
  no.~4, pp. 247--258, 2011. [Online]. Available:
  \url{https://doi.org/10.1007/s11416-011-0152-x}
\BIBentrySTDinterwordspacing

\bibitem{DBLP:conf/dimva/BruschiMM06}
\BIBentryALTinterwordspacing
D.~Bruschi, L.~Martignoni, and M.~Monga, ``Detecting self-mutating malware
  using control-flow graph matching,'' in \emph{Detection of Intrusions and
  Malware {\&} Vulnerability Assessment, Third International Conference,
  {DIMVA} 2006, Berlin, Germany, July 13-14, 2006, Proceedings}, ser. Lecture
  Notes in Computer Science, R.~B{\"{u}}schkes and P.~Laskov, Eds., vol.
  4064.\hskip 1em plus 0.5em minus 0.4em\relax Springer, 2006, pp. 129--143.
  [Online]. Available: \url{https://doi.org/10.1007/11790754\_8}
\BIBentrySTDinterwordspacing

\bibitem{DBLP:conf/cikm/ChaeHKKI13}
\BIBentryALTinterwordspacing
D.~Chae, J.~Ha, S.~Kim, B.~Kang, and E.~G. Im, ``Software plagiarism detection:
  a graph-based approach,'' in \emph{22nd {ACM} International Conference on
  Information and Knowledge Management, CIKM'13, San Francisco, CA, USA,
  October 27 - November 1, 2013}, Q.~He, A.~Iyengar, W.~Nejdl, J.~Pei, and
  R.~Rastogi, Eds.\hskip 1em plus 0.5em minus 0.4em\relax {ACM}, 2013, pp.
  1577--1580. [Online]. Available:
  \url{https://doi.org/10.1145/2505515.2507848}
\BIBentrySTDinterwordspacing

\bibitem{DBLP:conf/sec/SunZXMX14}
\BIBentryALTinterwordspacing
X.~Sun, Y.~Zhongyang, Z.~Xin, B.~Mao, and L.~Xie, ``Detecting code reuse in
  android applications using component-based control flow graph,'' in
  \emph{{ICT} Systems Security and Privacy Protection - 29th {IFIP} {TC} 11
  International Conference, {SEC} 2014, Marrakech, Morocco, June 2-4, 2014.
  Proceedings}, ser. {IFIP} Advances in Information and Communication
  Technology, N.~Cuppens{-}Boulahia, F.~Cuppens, S.~Jajodia, A.~A.~E. Kalam,
  and T.~Sans, Eds., vol. 428.\hskip 1em plus 0.5em minus 0.4em\relax Springer,
  2014, pp. 142--155. [Online]. Available:
  \url{https://doi.org/10.1007/978-3-642-55415-5\_12}
\BIBentrySTDinterwordspacing

\bibitem{DBLP:conf/iclr/GuoRLFT0ZDSFTDC21}
\BIBentryALTinterwordspacing
D.~Guo, S.~Ren, S.~Lu, Z.~Feng, D.~Tang, S.~Liu, L.~Zhou, N.~Duan,
  A.~Svyatkovskiy, S.~Fu, M.~Tufano, S.~K. Deng, C.~B. Clement, D.~Drain,
  N.~Sundaresan, J.~Yin, D.~Jiang, and M.~Zhou, ``Graphcodebert: Pre-training
  code representations with data flow,'' in \emph{9th International Conference
  on Learning Representations, {ICLR} 2021, Virtual Event, Austria, May 3-7,
  2021}.\hskip 1em plus 0.5em minus 0.4em\relax OpenReview.net, 2021. [Online].
  Available: \url{https://openreview.net/forum?id=jLoC4ez43PZ}
\BIBentrySTDinterwordspacing

\bibitem{DBLP:conf/mlsys/KaufmanPZM0SB21}
S.~J. Kaufman, P.~M. Phothilimthana, Y.~Zhou, C.~Mendis, S.~Roy, A.~Sabne, and
  M.~Burrows, ``A learned performance model for tensor processing units,'' in
  \emph{Proceedings of Machine Learning and Systems 2021, MLSys 2021, virtual,
  April 5-9, 2021}, A.~Smola, A.~Dimakis, and I.~Stoica, Eds.\hskip 1em plus
  0.5em minus 0.4em\relax mlsys.org, 2021.

\end{thebibliography}

\end{document}